\documentclass[12pt]{article}
\usepackage{multirow}
\usepackage{setspace, color, colortbl}
\usepackage{bm,amsmath,amsfonts}
\usepackage[round]{natbib}
\usepackage{subcaption}
\usepackage{makecell}
\usepackage{adjustbox}
\usepackage[utf8]{inputenc}
\usepackage{booktabs}
\usepackage[english]{babel}
\usepackage{algorithm,algpseudocode}
\usepackage{color,hyperref,hypernat}
\definecolor{darkblue}{rgb}{0.0,0.0,0.3}
\hypersetup{colorlinks,breaklinks,
            linkcolor=darkblue,urlcolor=darkblue,
            anchorcolor=darkblue,citecolor=darkblue}
\usepackage{mathtools}
\DeclarePairedDelimiter\abs{\lvert}{\rvert}

\usepackage{graphicx,psfrag,epsf}
\usepackage{enumerate, upgreek}
\usepackage{mathtools}
\usepackage{mathrsfs}
\definecolor{Gray}{gray}{0.9}
\newcommand{\code}[1]{\texttt{#1}}

\newtheorem{proposition}{Proposition}
\newtheorem{lemma}{Lemma}
\newtheorem{corollary}{Corollary}
\algnewcommand\algorithmicinput{\textbf{Update:}}
\algnewcommand\Update{\item[\algorithmicinput]}
\algnewcommand\Initialize{\item[{\textbf{Initialize:}}]}
\algnewcommand\Input{\item[{\textbf{Input:}}]}
\algnewcommand\Transform{\item[{\textbf{Transform:}}]}

\newcommand{\beginsupplement}{%
        \setcounter{table}{0}
        \renewcommand{\thetable}{S\arabic{table}}%
        \setcounter{figure}{0}
        \renewcommand{\thefigure}{S\arabic{figure}}%
     }

\makeatletter
\renewenvironment{titlepage}
 {%
  \if@twocolumn
    \@restonecoltrue\onecolumn
  \else
    \@restonecolfalse\newpage
  \fi
  \thispagestyle{empty}%
 }
 {%
  \if@restonecol
    \twocolumn
  \else
    \newpage
  \fi
 }
\makeatother

\begin{document}

\onehalfspace
\begin{titlepage}

\title{\textbf{The Reciprocal Bayesian LASSO}}
\author{Authors}
\author{Himel Mallick$^{1,*}$, Rahim Alhamzawi$^{2,3}$, Erina Paul$^{1}$, Vladimir Svetnik$^{1}$\ \\ \\
$^1$Biostatistics and Research Decision Sciences, Merck \& Co., Inc., Rahway, NJ, USA\\
$^2$Department of Statistics, University of Al-Qadisiyah, Al Diwaniyah, Iraq\\
$^3$Center for Scientific Research and Development, Nawroz University, Duhok, Iraq\\
$^*$Corresponding Author Email: \href{himel.mallick@merck.com}{himel.mallick@merck.com}}
\date{\today}
\maketitle
\thispagestyle{empty}
\begin{abstract}
\footnotesize
A reciprocal LASSO (rLASSO) regularization employs a decreasing penalty function as opposed to conventional penalization approaches that use increasing penalties on the coefficients, leading to stronger parsimony and superior model selection relative to traditional shrinkage methods. Here we consider a fully Bayesian formulation of the rLASSO problem, which is based on the observation that the rLASSO estimate for linear regression parameters can be interpreted as a Bayesian posterior mode estimate when the regression parameters are assigned independent inverse Laplace priors. Bayesian inference from this posterior is possible using an expanded hierarchy motivated by a scale mixture of double Pareto or truncated normal distributions. On simulated and real datasets, we show that the Bayesian formulation outperforms its classical cousin in estimation, prediction, and variable selection across a wide range of scenarios while offering the advantage of posterior inference. Finally, we discuss other variants of this new approach and provide a unified framework for variable selection using flexible reciprocal penalties. All methods described in this paper are publicly available as an R package at: \href{https://github.com/himelmallick/BayesRecipe}{https://github.com/himelmallick/BayesRecipe}.
\end{abstract}
\noindent KEYWORDS: Bayesian Regularization; Variable Selection; Reciprocal LASSO; Nonlocal Priors; MCMC; Penalized Regression
\end{titlepage}


\section{Introduction} \label{section1}

This paper concerns the development of a Bayesian analogue of the reciprocal LASSO (rLASSO, \citet{Song2015}) in a classical linear regression model ($\bm{y}=X\bm{\beta} + \bm{\epsilon}$) that results from the following regularization problem:
\begin{equation} \label{himel1}
Q(\bm{\beta}) = \underset{\bm{{\beta}} }{\text{min}} \ (\bm{y}-X\bm{\beta})'(\bm{y}-X\bm{\beta}) + \lambda\sum_{j=1}^{p}\frac{1}{|{\beta}_{j}|} I\{\beta_j \neq 0\},
\end{equation}
where $I(.)$ denotes an indicator function and $\lambda>0$ is the tuning parameter that controls the degree of penalization. Throughout the course of the paper, we assume that $\bm{y}$ and $X$ have been centered at 0 so there is no intercept in the model, where $\bm{y}$ is the $n \times 1$ vector of centered responses, X is the $n \times p$ matrix of standardized regressors, $\bm{\beta}$ is the $p \times 1$ vector of coefficients to be estimated, and $\bm{\epsilon}$ is the $n \times 1$ vector of independent and identically distributed normal errors with mean 0 and variance $\sigma^2$. \par 

Compared to traditional penalization functions that are usually symmetric about 0, continuous and nondecreasing in $(0, \infty)$, the rLASSO penalty functions are decreasing in $(0, \infty)$, discontinuous at 0, and converge to infinity when the coefficients approach zero. From a theoretical standpoint, rLASSO shares the same oracle property and same rate of estimation error with other LASSO-type penalty functions. An early reference to this class of models can be found in \citet{Song2015}, with more recent papers focusing on large sample asymptotics, along with computational strategies for frequentist estimation \citep{Song2018, Shin2018}. \par

Our approach differs from this line of work by adopting a Bayesian perspective on rLASSO estimation. Ideally, a Bayesian solution can be obtained by placing appropriate priors on the regression coefficients that will mimic the effects of the rLASSO penalty. As apparent from (\ref{himel1}), this arises in assuming a prior for $\bm{\beta}$ that decomposes as a product of independent inverse Laplace (double exponential) densities:
\begin{equation} \label{himel2}
\pi(\bm{\beta}) = \prod_{j=1}^{p}  \frac{\lambda}{2\beta_j^2}\exp\{-\frac{\lambda}{|\beta_j|}\}I\{\beta_j \neq 0\}.
\end{equation}
Rather than minimizing \eqref{himel1}, we solve the problem by constructing a Markov chain having the joint posterior for $\bm{\beta}$ as its stationary distribution with the minimizer of \eqref{himel1} as its global mode:
\begin{equation} \label{himel3}
\pi(\bm{\beta}|\bm{y}) = \exp\{-Q(\bm{\beta})\}.
\end{equation}
There are several motivations for undertaking a Bayesian approach to the rLASSO problem. First and foremost, the Bayesian construction offers a flexible framework endowed with richer model summaries, better performance in estimation and prediction, and more nuanced uncertainty quantification compared to the classical method. Second, the Bayesian rLASSO is computationally efficient, leading to scalable MCMC algorithms with good convergence and mixing properties. Third, the multimodal nature of the optimization problem \eqref{himel1} is one of the strongest arguments for pursuing a fully Bayesian approach, as summarizing a multimodal surface with a single frequentist point estimate can be vastly misleading \citep{Polson2014}. \par 

This multimodal phenomenon becomes especially apparent by taking a closer look at the classical rLASSO estimation, which relies on cross-validation to estimate the tuning parameter $\lambda$, followed by a computationally demanding Monte Carlo optimization procedure to search for the best model. From a practical standpoint, the algorithmic development of frequentist rLASSO has not been a priority in previous works, and to our knowledge, there are not many publicly available software tools that implement the rLASSO and no simple implementation relying on existing packages seems straightforward. \citet{Song2018} and \citet{Shin2018} report serious computational difficulties while attempting to minimize \eqref{himel1}, eventually falling back to approximate, non-cross-validated strategies with reduced computational burden for ultra-high-dimensional problems. Our sampling-based approaches, on the other hand, seem to be very effective at efficiently exploring the parameter space, offering a principled way of averaging over the uncertainty in the penalty parameter. The implementation of these methods with source code, documentation, and tutorial data are made freely available at the \code{BayesRecipe} software package: \href{https://github.com/himelmallick/BayesRecipe}{https://github.com/himelmallick/BayesRecipe}. \par 

In summary, the major contributions of the paper are as follows: (i) introduction of the rLASSO prior as a shrinkage prior in Bayesian analysis, (ii) a set of data augmentation strategies motivated by a novel scale mixture representation of the rLASSO density, (iii) scalable MCMC algorithms for posterior inference, (iv) an informed software implementation, and (v) extensibility of the method to general models as well as to other penalties, greatly expanding the scope of reciprocal regularization beyond linear regression. \par

The remainder of the paper is organized as follows. In Section \ref{section2}, we describe the rLASSO  prior and study its various properties. In Section \ref{section3}, we develop our Bayesian rLASSO estimator. A detailed description of the MCMC sampling scheme is laid out in Section \ref{section4}. Empirical evidence of the attractiveness of the method is demonstrated in Section \ref{section5} via extensive simulation studies and real data analyses. Finally, in Section \ref{section6}, we discuss other variants of this new approach and provide a unified framework for variable selection using flexible reciprocal penalties. We conclude with further discussion in this area in Section \ref{section7}. Proofs and derivations (referenced in Section \ref{section2}), algorithmic details (referenced in Section \ref{section4}), and additional extensions (referenced in Section \ref{section6}) are provided in the Appendix.


\section{Prior Elicitation} \label{section2}
\subsection{Connection with other priors}
Without loss of generality, we consider a one-dimensional rLASSO prior as follows:
\begin{equation} \label{himel4}
\pi(\beta) = \frac{\lambda}{2\beta^2}\exp\{-\frac{\lambda}{|\beta|}\}I\{\beta\neq 0\},
\end{equation}
where $\lambda>0$ is a scale parameter. We refer to this distribution as Inverse Double Exponential (IDE) distribution, which has Cauchy-like tails and no first- and second-order moments \citep{Woo2009}.

\citet{Song2015} noted that the rLASSO prior belongs to the class of nonlocal priors (NLPs) \citep{Johnson2010, Johnson2012}, sharing a very similar nonlocal kernel with the piMOM prior (differing only in the power of $\beta$ in the exponential component). In sharp contrast to most popular shrinkage priors that assign a non-zero probability near zero, nonlocal priors are exactly zero whenever a model parameter approaches its null value (i.e. $\beta=0$). Relative to local priors (LPs), NLPs discard spurious covariates faster as the sample size $n$ grows, while preserving exponential learning rates to detect non-zero coefficients \citep{Rossell2017}. This is particularly apparent from the heavy tails of NLPs, which is appealing in avoiding over-shrinkage away from the origin. \par

\begin{figure}
\begin{center}
\includegraphics[width=0.65\textwidth]{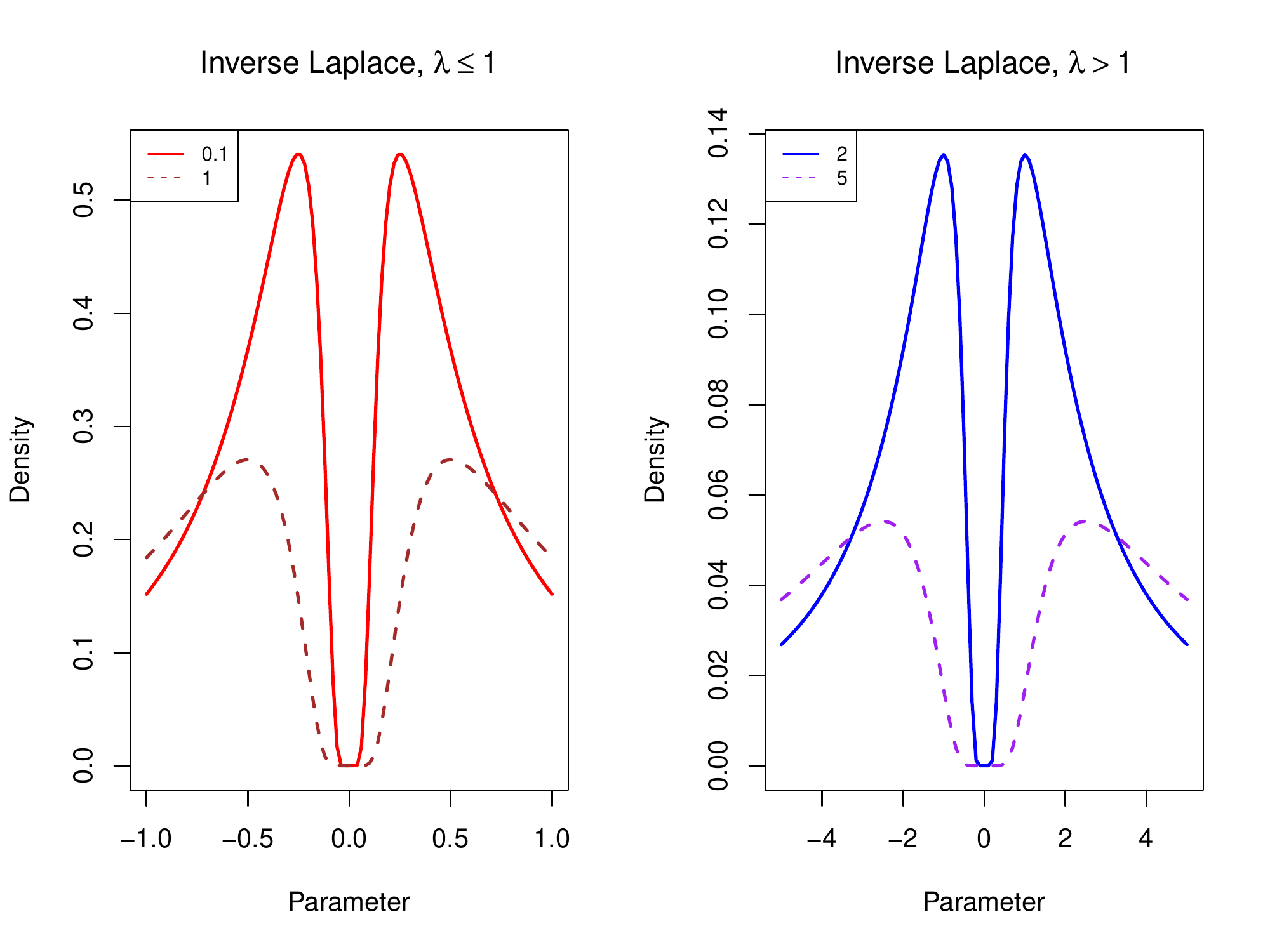}
\caption{The marginal densities of the rLASSO prior for a single regression coefficient for varying $\lambda$. A smaller $\lambda$ leads to a tighter interval near zero (left), which eventually widens as $\lambda$ increases (right).}\label{figure1}
\end{center}
\end{figure}

From Figure \ref{figure1}, it is clear that the hyperparameter $\lambda$ represents a scale parameter that determines the dispersion of the prior around 0. Therefore, in order to facilitate sparse recovery, $\lambda$ should be relatively small. This fact is somewhat counter-intuitive given that most LASSO-type regularization methods typically impose a large value of $\lambda$ to penalize coefficients. As highlighted by an increasing body of literature, Bayesian variable selection procedures based on nonlocal priors have been shown to outperform other popular variable selection methods in a wide range of applications \citep{Nikooienejad2016, Sanyal2019, Nikooienejad2020}, leading to superior posterior consistency properties in both high-dimensional and ultra-high-dimensional settings \citep{Rossell2017, Shin2018}. \par


Existing methods for nonlocal priors mostly focus on Laplace approximation of the marginal likelihood \citep{Johnson2010, Johnson2012, Shin2018}, which is known to be highly inaccurate in high-dimensional or small sample size problems \citep{Ruli2016}. Additionally, in the presence of increased model complexity, Laplace’s method can be unstable as the numerical integration may eventually require a large number of quadrature points, leading to a higher computational overhead \citep{Ruli2016}. Furthermore, existing stochastic search algorithms such as the Simplified Shotgun Stochastic Search with Screening (S5) algorithm proposed by \citet{Shin2018}, lack theoretical guarantee of convergence and may lead to strongly biased Monte Carlo estimates of quantities of interest, despite facilitating scalable computation for model selection \citep{Hahn2015}. Finally, due to the non-propriety of NLPs, simple posterior sampling from these priors has been historically difficult, which is appealing from a practitioner's perspective. \citet{Rossell2017} recently proposed hierarchical NLPs with some similarities to the rLASSO prior, but they lack a one-to-one correspondence with an equivalent frequentist estimator that facilitates comparative study of maximum efficiency properties in complementary Bayesian-frequentist domains (much as has been the case for LASSO-type penalties and LPs over the last decade or so).\par

Within the broader class of Bayesian regularized estimators in high-dimensional regression, there has been widespread interest in cases where the implied prior corresponds to a scale mixture of normal (SMN) \citep{Kyung2010}. Many estimators in this class share the favorable regularization property (e.g., heavy tails) of the Bayesian rLASSO model. In virtually all of these models, the primary difficulty is the induction of suitable latent variables to make the corresponding MCMC sampling convenient with tractable conditional posteriors. However, due to the lack of a similar closed-form SMN representation of the rLASSO density, direct posterior sampling from \eqref{himel3} is complicated. In light of this, we introduce a novel characterization of the rLASSO prior as a double Pareto scale mixture that yields proper posteriors and leads to efficient MCMC algorithms, as we describe in the sequel.

\subsection{Scale Mixture of Double Pareto (SMDP) Representation}

As shorthand notation, let $\beta \sim $ IDE($\lambda)$ denote that $\beta$ has density \eqref{himel4} which can be represented as a scale mixture of double Pareto (or inverse uniform) densities leading to computational simplifications, as detailed below.

\begin{proposition} Let $\beta \sim \text{Double Pareto} \ (\eta, 1)$ and $\eta \sim \text{Inverse Gamma} \ (2, \lambda)$, where $\eta>0$. The resulting marginal density for $\beta$ is IDE ($\lambda)$.
\end{proposition}

 \subsection{Scale Mixture of Truncated Normal (SMTN) Representation}
Noting that the double Pareto distribution can be further decomposed as a truncated scale mixture of Laplace or normal distributions \citep{Armagan2013a}, we have a second representation as follows:

\begin{proposition} Let $\beta \sim N(0, \tau)I(|\beta| > \eta)$, $\tau \sim Exp(\zeta^2/2)$, $\zeta \sim Exp (\eta)$, and \\ $\eta \sim \text{Inverse Gamma}(2, \lambda)$, where $\lambda>0$. The resulting marginal density for $\beta$ is IDE($\lambda)$.
\end{proposition}

The proofs of Propositions 1 and 2 are deferred to the Appendix.


\section{Model Hierarchy and Prior Distributions} \label{section3}
\subsection{Exploiting the SMDP Representation}\label{section3A}
Based on Proposition 1, assuming a non-informative scale-invariant marginal prior on $\sigma^2$, i.e. $\pi(\sigma^2) \propto 1 \slash\sigma^2$ and transforming $u = \frac{1}{\eta}$, we have the following expanded hierarchy:
$$
\bm{y}^{n \times 1}|X, \bm{\beta}, \sigma^2 \sim N_n(X\bm{\beta}, \sigma^2 I_{n}),
$$
$$
\bm{\beta}^{p \times 1}|\bm{u} \sim \displaystyle \prod_{j=1}^{p} \frac{1}{\text{Uniform} (-u_j, u_j)},\\ 
$$
$$
\bm{u}^{p \times 1}|\lambda \sim \displaystyle \prod_{j=1}^{p} \text{Gamma} (2,\lambda),\\
$$
$$
\sigma^2 \sim \pi(\sigma^2).
$$
\subsection{Exploiting the SMTN Representation}
Similarly, by virtue of Proposition 2, letting $\beta_j|\sigma \sim \text{IDE} \ (\lambda\sigma)$ independently for $j = 1, \ldots, p$, we have the following hierarchical formulation:
$$
\bm{y}^{n \times 1}|X, \bm{\beta}, \sigma^2 \sim N_n(X\bm{\beta}, \sigma^2 I_{n}),
$$
$$
\bm{\beta}^{p \times 1}|\bm{\tau}, \bm{u}, \sigma^2 \sim \displaystyle \prod_{j=1}^{p} N(0, \sigma^2 \tau_{j}^2) I\{|\beta_j|> \frac{\sigma}{u_j}\},\\ 
$$
$$
\bm{\tau}^{p \times 1}|\bm{\zeta} \sim \displaystyle \prod_{j=1}^{p} \text{Exp} (\zeta_{j}^2/2),\\
$$
$$
\bm{\zeta}^{p \times 1}|\bm{u} \sim \displaystyle \prod_{j=1}^{p} \text{Exp} (\frac{1}{u_j}),\\
$$
$$
\bm{u}^{p \times 1}|\lambda \sim \displaystyle \prod_{j=1}^{p} \text{Gamma} (2,\lambda),\\
$$
$$
\sigma^2 \sim \pi(\sigma^2).
$$
\subsection{Connection between regular and reciprocal Bayesian LASSOs}
Proposition 1 reveals an interesting contrast between two distant cousins: regular Bayesian LASSO and reciprocal Bayesian LASSO, in light of their respective hierarchical formulations. To observe that, we rewrite the Bayesian LASSO hierarchical model using the scale mixture of uniform (SMU) representation of the Laplace density \citep{Mallick2014}:
$$
\bm{y}^{n \times 1}|X, \bm{\beta}, \sigma^2 \sim N_n(X\bm{\beta}, \sigma^2 I_{n}),
$$
$$
\bm{\beta}^{p \times 1}|\bm{u} \sim \displaystyle \prod_{j=1}^{p} \text{Uniform} (-u_j ,u_j),\\ 
$$
$$
\bm{u}^{p \times 1}|\lambda \sim \displaystyle \prod_{j=1}^{p} \text{Gamma} (2,\lambda),\\
$$
$$
\sigma^2 \sim \pi(\sigma^2).
$$
Return now to the reciprocal Bayesian LASSO model in Section \ref{section3A}, which indicates that both the tuning parameter and the latent variables induce an exact opposite effect on $\bm{\beta}$, as expected. In particular, the reciprocal Bayesian LASSO demands a small $\lambda$ for optimal performance, whereas a large value of $\lambda$ is desirable for the Bayesian LASSO. Second, as evident from the respective uniform and inverse uniform conditional priors, the LASSO density assigns a high prior probability near zero, in stark contrast to the rLASSO prior which assigns a high probability only when away from zero. As a result of these differences, the resulting posteriors are also remarkably different, which correspond to doubly truncated and internally truncated multivariate normal distributions, respectively for regular and reciprocal Bayesian LASSOs. We derive the Gibbs sampler using the SMDP representation (described in Section 4.1) only to reveal a close connection with the Bayesian LASSO and do not use it in our experiments. 

\section{Computation for the Reciprocal Bayesian LASSO} \label{section4}
\subsection{Full Posterior Distributions}
The full conditional posterior distributions can be derived using simple algebra for the SMDP prior specification:
$$
\bm{\beta}|\bm{y},X,\bm{u},\lambda,\sigma^2 \sim  N_p(\hat {\bm{\beta}}_{\text{MLE}},\sigma^2 {(X'X)}^{-1} ) \displaystyle \prod_{j=1}^{p} I\{|\beta_{j}| > \frac{1}{u_j}\},
$$
$$
\bm{u}|\bm{y},X,\bm{\beta},\lambda,\sigma^2 \sim  \displaystyle \prod_{j=1}^{p} \text{Exp}(\lambda)  I\{u_j>\frac{1}{|\beta_{j}|}\},
$$
$$
\sigma^2|\bm{y},X,\bm{\beta},\bm{u},\lambda \sim \text{Inverse-Gamma}(\frac{n-1}{2}, \frac{1}{2}(\bm{y}-X\bm{\beta})'(\bm{y}-X\bm{\beta})).
$$

Similarly, we obtain a simple data augmentation Gibbs sampler using the SMTN representation as follows:
$$
\bm{\beta}|\bm{y},X,\bm{\tau}, \bm{\zeta}, \bm{u}, \lambda,\sigma^2 \sim N_p((X'X + T^{-1})^{-1} X^{'}\bm{y},\sigma^2 {(X'X + T^{-1})}^{-1}) \displaystyle \prod_{j=1}^{p} I\{|\beta_{j}| > \frac{\sigma}{u_j}\},
$$
$$
\bm{\tau^{-1}}|\bm{y},X,\bm{\beta}, \bm{\zeta}, \bm{u}, \lambda,\sigma^2 \sim  \displaystyle \prod_{j=1}^{p} \text{Inverse-Gaussian}(\sqrt{\frac{\zeta_j^{2}\sigma^2}{\beta_j^{2}}}, \zeta_j^{2}),
$$
$$
\bm{\zeta}|\bm{y},X,\bm{\beta}, \bm{\tau}, \bm{u}, \lambda,\sigma^2 \sim  \displaystyle \prod_{j=1}^{p} \text{Gamma}(2, \left (\frac{|\beta_j|}{\sigma} + \frac{1}{u_j} \right )),
$$
$$
\bm{u}|\bm{y},X,\bm{\beta}, \bm{\zeta}, \bm{\lambda}, \lambda,\sigma^2 \sim  \displaystyle \prod_{j=1}^{p} \text{Exp}(\lambda)  I\{u_j>\frac{\sigma}{|\beta_{j}|}\},
$$
$$
\sigma^2|\bm{y},X,\bm{\beta}, \bm{\zeta}, \bm{\lambda}, \bm{u},\lambda \sim \text{Inverse-Gamma}(\frac{n-1+p}{2}, \frac{R + \bm{\beta}^{'}T^{-1}\bm{\beta}}{2}) I\{\sigma^2 < \text{Min}_j(\beta_j^2 u_j^2)\},
$$\\
where $R = (\bm{y}-X\bm{\beta})'(\bm{y}-X\bm{\beta})$ and $T = \text{diag} (\tau_1, \dots, \tau_p)$. All the resulting conditionals are standard, making them easy to implement using existing sampling algorithms with the exception of $\bm{\beta}|.$, which follows a mid-truncated multivariate normal distribution \citep{Kim2007}, for which, we resort to an efficient sampling technique developed by \citet{Rossell2017}. In our experience, the resulting Gibbs samplers are efficient with fast rates of convergence and mixing. We document the algorithmic details of these two Bayesian samplers in the Appendix (Figures \ref{figureS1}-\ref{figureS2}).

\subsection{Choosing the Reciprocal Bayesian LASSO Hyperparameter}
An important aspect of the reciprocal Bayesian LASSO implementation is the careful selection of the hyperparameter $\lambda$, which critically impacts downstream posterior inference. While there are several ways to choose $\lambda$, we propose three well-known procedures, drawing upon a vast array of literature from LPs and NLPs. In what follows, we refer to the resulting methods as BayesA (Apriori Estimation), BayesB (Empirical Bayes), and BayesC (MCMC), respectively. 

\subsubsection{Apriori Estimation}
In many cases, the tuning parameter can be fixed ahead of time to reflect a particular desired shape of the rLASSO penalty function. To this end, we extend the procedure of \citet{Shin2018} for NLPs to the rLASSO prior and select $\lambda$ such that the $L_1$ distance between the posterior distribution of the regression parameters under the null distribution (i.e. $\beta=0$) and the rLASSO prior distribution on these parameters is constrained to be less than a specified value (e.g., $\frac{1}{\sqrt{p}}$). By choosing an optimal $\lambda$ so that the intersection of these two null distributions falls below a specified threshold, this procedure approximately bounds the probability of false positives in the model, while maintaining sensitivity to detect large effects \citep{Nikooienejad2016}. For brevity, we skip the technical details of the algorithm and refer the readers to \citet{Shin2018} and references therein.

\subsubsection{Empirical Bayes by Marginal Maximum Likelihood}
From a practical perspective, it is desirable to select $\lambda$ by adaptively learning from the observed sparsity or signal level in a dataset. Following \citet{Park2008},  we implement a Monte Carlo EM algorithm that complements a Gibbs sampler by essentially treating $\lambda$ as ``missing data'' and then iteratively updates $\lambda$ by maximizing the marginal likelihood. Using a similar derivation in \citet{Park2008}, we can carry out this update using a closed-form expression 
$$
\lambda^{(k)} = \frac{2p}{\sum_{j=1}^p E_{\lambda^{(k-1)} }(u_j)},$$ 
which corresponds to the maximizer of the expected value of the `complete-data' log-likelihood
$$
Q(\lambda | \lambda^{(k)}) =  p \ln(\lambda^2) - \lambda \sum_{j=1}^p E_{\lambda^{(k-1)} }(u_j) + \text{terms not involving $\lambda$},
$$
where the conditional expectations are just the posterior expectations under the hyperparameter $\lambda^{(k-1)}$ (the estimate from iteration $k-1$), and therefore they can be estimated using the sample averages from a single run of the corresponding Gibbs sampler. 
\subsubsection{Hyperpriors for the rLASSO Parameter}
As an alternative to choosing $\lambda$ explicitly, we can also update $\lambda$ by assigning a diffuse conjugate hyperprior. From \eqref{himel4}, we observe that the posterior for $\lambda$ given $\bm{\beta}$ is conditionally independent of $\bm{y}$ and takes the form
$$ 
\pi(\lambda|\bm{\beta}) \propto \lambda^{2p} \exp\{-\lambda \displaystyle \sum_{j=1}^{p} \frac{1}{|\beta_{j}|}\}\pi(\lambda).
$$\par Therefore, assuming a $\text{Gamma}(a,b)$ prior on $\lambda$, it can be updated along with other parameters in the model by generating samples from $\text{Gamma}(a+2p, b+ \displaystyle \sum_{j=1}^{p} \frac{1}{|\beta_{j}|})$. \par

\subsection{Posthoc Variable Selection}
Similar to existing Bayesian regularization methods, a full posterior exploration of $\bm{\beta}$ does not automatically induce sparsity, and therefore, variable selection must be conducted in a posthoc manner by sparsification of posterior summaries \citep{Hahn2015}. Here we consider a hybrid Bayesian-frequentist strategy that achieves sparse selection by backpropagating the Bayesian estimate of the tuning parameter $\lambda$ (viz. posterior median) in the optimization problem \eqref{himel1} to solve a frequentist reciprocal LASSO problem. This approach has recently been considered by \citet{Leng2014} in the context of Bayesian adaptive LASSO regression, which led to surprising improvement in variable selection performance over published methods. As the name implies, this procedure is inspired by the use of backpropagation in neural network models, which we refer to as Frequentist Backpropagation (FBP).

\section{Numerical Studies} \label{section5}
\subsection{Simulation Results}

Our main interest in this section is to study the advantages and disadvantages of the Bayesian reciprocal LASSO in simulation experiments. There has been an enormous amount of work in the statistics and machine learning community dealing with regularization and feature selection in a wide spectrum of problems \citep{Kyung2010, Tibshirani2011, Mallick2013, Vidaurre2013, Song2018, Emmert-Streib2019, Hastie2019, Bhadra2019, Van-Erp2019, Bai2020}. While an exhaustive benchmarking is beyond our scope, we restrict our focus on evaluating reciprocal LASSO and related methods for two primary reasons. First, previous studies have extensively compared rLASSO to other published methods in diversified and realistic scenarios, justifying the need to avoid unnecessary duplication of effort \citep{Shin2018, Song2015, Song2018}. Second, our restricted scope enables head-to-head comparison of the frequentist and Bayesian machineries of reciprocal regularization, which is particularly useful when rLASSO has an edge over other competing methods and the Bayesian underpinning of the problem can potentially enhance performance through further refinement. \par

In the absence of a `gold standard' publicly available routine for computing the rLASSO solution path, we explored several candidate algorithms, ultimately settling on the S5 proposal of \citet{Shin2018} as the most straightforward to implement. Although this may appear conceptually modest, the implementation involves additional programming efforts beyond those encountered in established workflows. Our simulation study thus adds enormous practical value to the extant literature by serving both frequentist and Bayesian implementations of reciprocal LASSO for practitioners (all methods and experiments described in this paper are freely available online at: \url{https://github.com/himelmallick/BayesRecipe}). A detailed description of the S5 algorithm, which has been omitted here, can be found in \citet{Shin2018}. 
\par

Briefly, S5 is a stochastic search method that screens candidate covariates at each step, exploring regions of high posterior model probability based on the rLASSO objective function \eqref{himel1}, where screening is defined based on the correlation between the residuals of the regression using the current model and the remaining covariates. In our implementation, 30 iterations are used within each temperature, with other relevant  parameters fixed at default values recommended by the authors \citep{Shin2018}. Following \citet{Shin2018}, we choose the rLASSO hyperparameter $\lambda$ using the apriori estimation procedure described in Section 4.2.1. A least squares refitting (debiasing) step is carried out on the selected model to estimate the final model coefficients. \par

For the Bayesian reciprocal LASSO, we implement the Gibbs sampler described in Section 4 using the SMTN representation. We set $a=b=0.001$ when estimating $\lambda$ (except for the apriori estimation method, where $\lambda$ is estimated prior to Gibbs sampling). We run the corresponding Gibbs samplers for $11,000$ iterations, discarding the first $1,000$ as burn-in. This choice of running parameters appear to work satisfactorily based on the convergence diagnostics. We use the posterior mean as our point estimator and deploy the FBP strategy described in Section 4.2 for variable selection. \par

For the competing Bayesian methods, we use the horseshoe regression \citep{Carvalho2010}, as implemented in the R package \code{horseshoe}, which employs the algorithm proposed by \citet{Bhattacharya2016}.  For variable selection using the horseshoe regularization, we consider two posthoc sparsification procedures: 1) the ``decoupled shrinkage and selection" (DSS) method proposed by \citet{Hahn2015}, and (2) the 95$\%$ credible interval, as advocated by \citet{Park2008}. In our experience, these approaches lead to very similar results, with the former requiring additional tuning (e.g., the regularization parameter in the surrogate optimization problem in DSS). As a result, to simplify narrative, we report variable selection results related to the $95\%$ equal-tailed credible intervals for the horseshoe estimator. As before, we run the corresponding horseshoe slice sampler for $11,000$ iterations, discarding the first $1,000$ as burn-in, with other parameters fixed at their default values.\par

We simulate data from the true model $\bm{y}=X\bm{\beta_0} +\bm{\epsilon}, \ \  \bm{\epsilon} \sim  \text{N}  (\bm{0},\sigma^2I_n)$, and consider both $n \leq p$ and $n>p$ settings as well as a range of sparse and dense models with diverse effect sizes and collinearity patterns (Table \ref{table1}). The design matrix $X$ is generated from the multivariate normal distribution $N (\bm{0}, \Sigma)$, where $\Sigma$ has one of the following covariance structures for varying correlation strength ($\rho$):

\begin{enumerate}
\item Case I (IS): Isotropic design, where $\Sigma = I_p$.
\item Case II (CS): Compound symmetry design, where $\Sigma_{ij} = \rho$ whenever $i \neq j$ and $\Sigma_{ii} = 1$, for all $1 \leq i \leq j \leq p$.
\item Case III (AR): Autoregressive correlated design, where $\Sigma_{ij} = \rho^{|i-j|}$ for all $1 \leq i \leq j \leq p$.
\end{enumerate}

\noindent For each parameter combination, we generate 100 datasets and each synthetic dataset is further partitioned into a training set and a test set. 
For performance measures, we compute the out-of-sample mean squared error (MSE) and the balanced accuracy rate \citep{Brodersen2010}, averaged over 100 simulation runs. The Balanced Accuracy Rate (BAR) is a comprehensive performance metric that combines both the sensitivity and specificity of a classifier, defined as 
$$ 
\text{BAR} = \frac{1}{2} \left ( \frac{\text{TP}}{\text{TP} + \text{FN}} + \frac{\text{TN}}{\text{TN} + \text{FP}} \right ),
$$
where TP, TN, FP, and FN denote the number of true positives, true negatives, false positives, and false negatives, respectively.  \par

These synthetic experiments reveal that the Bayesian hierarchical rLASSOs perform as well as or better than frequentist rLASSO in most of the examples, which is consistent across varying sparsity levels and signal strengths (Table \ref{table2}). Specifically when the underlying true model is sparse, the Bayesian methods perform better in terms of all measures with rLASSO comparing favorably to BayesA but less favorably to BayesB and BayesC in terms of MSE. This can be explained by the fact that compared to BayesB and BayesC, BayesA estimates the tuning parameter prior to MCMC sampling and fails to capture the model averaging effect of estimating $\lambda$, leading to larger estimation error. Overall, the BAR values for the Bayesian methods are all higher than frequentist rLASSO in sparse settings, meaning that the Bayesian methods can identify the true model more precisely. All the methods perform worse as the true model becomes dense. This is not surprising given that rLASSO assigns large penalties to small coefficients and avoids selecting overly dense models, yielding sub-optimal solutions in highly dense settings. The Bayesian methods closely follow this behavior, as not much variance is explained by introducing the prior when the underlying true model is not overly complicated. \par

In addition to the rLASSO-centric comparisons above, we note that the reciprocal Bayesian LASSO has the smallest MSE in 13 out of 24 simulation setups and the largest BAR in 14 out of 24 scenarios, often performing as well as or better than the horseshoe method. Based on our numerical experiments, the highly dense models often lead to poor estimation for the horseshoe prior, although it remains the best method in the highly sparse situations, as expected. The reciprocal Bayesian LASSO, although often overshadowed in these highly sparse settings, remains competitive across a range of models as the overall sparsity level decreases. This suggests that consistent with the existing theory, the horseshoe method remains near optimal in extremely sparse settings and the reciprocal Bayesian LASSO is able to reduce this inevitable performance gap in less extreme settings with respect to either prediction accuracy or variable selection consistency or both in most of the examples.


\begin{table}[!h]
\begin{center}
\caption{Benchmarking configurations for the simulation study.} \label{table1}
\scalebox{0.9}{
\begin{tabular}{c c | c | c c c c | c}
\hline
\textbf{Model} & & \textbf{Setting} & $\bm{(n, p)}$ & $\bm{\Sigma}$&  $\bm{\rho}$ &  $\bm{\sigma}$ & $\bm{\beta_0}$\\ \hline
& (I) & & (50, 20) & IS & 0.0 & 3 &\\ 
& (II) & $n > p$ & (100, 10) & CS & 0.5 & 3 &\\ 
Highly Sparse & (III) & & (100, 50)& AR & 0.95 & 1.5 & $(5, 0, \ldots, 0)^{T}$\\ \cmidrule{3-7} 
& (IV) & & (50, 50)& IS & 0.0 & 1.5 & \citep{Tibshirani1996}\\
& (V) & $n \leq p $ & (100, 200)& CS & 0.5 & 3 &\\ 
& (VI) & & (50, 100)& AR & 0.95 & 1.5 &\\ \hline
& (VII) & & (400, 20) & IS & 0.0 & 3 &\\ 
& (VIII) & $n > p$ & (50, 20) & CS & 0.5 & 3 &\\ 
Fairly Sparse & (IX) & & (100, 10)& AR & 0.95 & 3 & $(3,1.5,0,0,2,0, \ldots, 0)^{T}$\\ \cmidrule{3-7} 
& (X) & & (100, 100)& IS & 0.0 & 1.5 & \citep{Tibshirani1996}\\
& (XI) & $n \leq p $ & (50, 50)& CS & 0.5 & 1.5 &\\ 
& (XII) & & (100, 200)& AR & 0.95 & 1.5 &\\ \hline
& (XIII) & & (400, 50) & IS & 0.0 & 3 &\\ 
& (XIV) & $n > p$ & (400, 200) & CS & 0.5 & 1.5 &\\ 
Moderately Sparse & (XV) & & (100, 50)& AR & 0.95 & 1.5 & $\pm (\frac{1}{2},\frac{3}{4},1,\frac{5}{4},\frac{3}{2}, 0, \ldots, 0)^{T}$\\ \cmidrule{3-7}  
& (XVI) && (50, 50)& IS & 0.0 & 1.5 & \citep{Shin2018}\\
& (XVII) & $n \leq p $ & (100, 200)& CS & 0.5 & 3 &\\ 
& (XVIII) & & (50, 200)& AR & 0.95 & 3 &\\ \hline
& (XIX) & & (50, 20) & IS & 0.0 & 3 &\\ 
&(XX) &  $n > p$ & (400, 100) & CS & 0.5 & 1.5 &\\ 
Highly Dense & (XXI) & & (400, 200)& AR & 0.95 & 3 & $(0.85, \ldots, 0.85)^{T}$\\ \cmidrule{3-7} 
& (XXII) & & (100, 200)& IS & 0.0 & 1.5 & \citep{Tibshirani1996}\\
& (XXIII) & $n \leq p $ & (100, 200)& CS & 0.5 & 3 &\\ 
& (XXIV) & & (50, 50)& AR & 0.95 & 1.5 &\\ \hline
\end{tabular}}
\end{center}
\end{table}


\begin{table}
\begin{center}
        \caption{Comparison of Bayesian and frequentist rLASSO methods as well as the horseshoe method by \citet{Carvalho2010}. Values are median out-of-sample mean squared error (MSE) and median percentage of correct selections as measured by Balanced Accuracy Rate (BAR) ($\text{Correct}\%$), summarized over 100 simulation runs. For each simulation scenario, the best method ($\text{Winner}$) is also reported. The bold numbers are the minimum MSE or the maximum BAR per scenario. BRL stands for Bayesian Reciprocal LASSO, whereas, HS is an abbreviation for the horseshoe method. For simplicity, we declare BRL as the winner in the event of a tie involving any BRL method.}\label{table2}
\scalebox{0.6}{
        \begin{tabular}{lrrrrrrrrrrll}\toprule
          &\multicolumn{2}{c}{\textbf{BayesA}}&\multicolumn{2}{c}{\textbf{BayesB}}&\multicolumn{2}{c}{\textbf{BayesC}}&\multicolumn{2}{c}{\textbf{rLASSO}}&\multicolumn{2}{c}{\textbf{Horseshoe}}&\multicolumn{2}{c}{\textbf{Winner}}
            \\\cmidrule(r){2-3}\cmidrule(r){4-5}\cmidrule(r){6-7}\cmidrule(r){8-9}  \cmidrule(r){10-11}\cmidrule(r){12-13} 
              \textbf{Model} & MSE & $\text{Correct}\%$ & MSE & $\text{Correct}\%$ & MSE & $\text{Correct}\%$ & MSE & $\text{Correct}\%$& MSE & $\text{Correct}\%$& MSE & $\text{Correct}\%$\\\midrule
              Highly Sparse (I) & 135.44 & \textbf{60.53} & 131.92 & 55.26 & \textbf{131.88} & 55.26 & 135.16 & 57.89 &132.04&\textbf{60.53}& BRL & BRL\\ 
  Highly Sparse (II) & 90.92 & \textbf{61.11} & 89.73 & 55.56 & \textbf{89.49} & 55.56 & 90.92 & 55.56 & 90.73 &\textbf{61.11}&BRL&BRL\\ 
  Highly Sparse (III) & 8.98 & 77.04 & \textbf{8.82} & 68.37 & \textbf{8.82} & 68.37 & 9.07 & 73.47 & 9.84 &\textbf{93.88}&BRL&HS\\ 
  Highly Sparse (IV) & 21.71 & 80.61 & 25.16 & 69.39 & 25.52 & 69.39 & 27.33 & 74.49 & \textbf{8.00} &\textbf{95.92}&HS & HS\\ 
  Highly Sparse (V) & 209.25 & 92.71 & 209.65 & 92.71 & 209.65 & 92.71 & 209.81 & 92.71 & \textbf{149.46} &\textbf{97.74}& HS &HS\\ 
  Highly Sparse (VI) & 16.03 & 90.91 & 17.43 & 82.32 & 17.45 & 81.82 & 17.86 & 87.37 & \textbf{8.92} &\textbf{99.49}&HS & HS\\ \hline
  Fairly Sparse (VII) & 84.46 & 60.29 & \textbf{82.83} & \textbf{94.12} & 83.17 & \textbf{94.12} & 84.46 & 61.76 & 84.39 &58.82 & BRL & BRL\\ 
  Fairly Sparse (VIII) & 131.02 & \textbf{58.82} & \textbf{129.19} & 55.88 & \textbf{129.19} & 55.88 & 131.21 & 55.88 & 129.70 &55.88&BRL&BRL\\ 
  Fairly Sparse (IX) & 89.90 & \textbf{57.14} & \textbf{89.47} & \textbf{57.14} & \textbf{89.47} & \textbf{57.14} & 89.90 & 53.57& 90.48 &52.38& BRL&BRL\\ 
  Fairly Sparse (X) & 13.62 & 86.60 & 12.73 & 86.08 & 12.64 & 86.08 & 13.56 & 86.08 & \textbf{7.96} &\textbf{95.36}&HS & HS\\ 
  Fairly Sparse (XI) & 20.50 & 80.85 & 22.91 & 71.28 & 22.78 & 71.28 & 26.11 & 74.47 & \textbf{9.98} &\textbf{94.68}&HS & HS\\ 
  Fairly Sparse (XII) & 12.52 & \textbf{76.99} & 12.03 & 76.23 &\textbf{ 11.88} & 76.23 & 12.67 & 76.23& 15.08&66.67& BRL&BRL\\ \hline
  Moderately Sparse (XIII) & 90.87 & 72.22 & 86.90 & \textbf{74.44} & \textbf{86.78} & 74.44\textbf{} & 90.87 & 72.22 & 90.97 &60.00&BRL&BRL\\ 
  Moderately Sparse (XIV) & 6.61 & 93.59 & 6.40 & \textbf{94.10} &\textbf{ 6.39} &\textbf{ 94.10} & 6.63 & 93.59 &7.03 &	91.03 & BRL & BRL\\ 
  Moderately Sparse (XV) & 9.01 & 55.56 & \textbf{8.75} & 55.56 & 8.76 & 54.44 & 9.10 & 52.22 &10.81 &\textbf{61.11}& BRL & HS\\ 
  Moderately Sparse (XVI) & 29.11 & 65.56 & 25.93 & 64.44 & 26.60 & 64.44 & 30.39 & 64.44 & \textbf{9.75} &\textbf{76.11}&HS &HS\\ 
  Moderately Sparse (XVII) & 208.73 & \textbf{52.56} & 207.13 & \textbf{52.56} & 207.13 & \textbf{52.56} & 209.07 & \textbf{52.56} & \textbf{146.66} & 48.46 & HS & BRL\\ 
  Moderately Sparse(XVIII) & 214.26 & \textbf{53.08} & 207.04 & 52.56 & 207.51 & 52.56 & 216.80 & 52.56 & \textbf{108.51} & 50.00 & HS & BRL\\ \hline
  Highly Dense (XIX) & 134.08 & 85.00 & 129.02 & 70.00 & \textbf{128.49 }& 62.50 & 134.20 & \textbf{90.00} & 132.67 & - & BRL & rLASSO\\ 
  Highly Dense (XX) & 97.74 & \textbf{30.00} & 97.00 & \textbf{30.00} & 97.00 & \textbf{30.00} & 97.80 & \textbf{30.00} &\textbf{8.55} & - & HS & BRL\\ 
  Highly Dense (XXI) & 206.53 & \textbf{15.00} & 212.85 & \textbf{15.00} & 212.85 & \textbf{15.00} & 217.20 & \textbf{15.00} &  \textbf{168.74} & -& HS & BRL\\ 
  Highly Dense (XXII) & 232.54 & \textbf{15.00} & 232.57 & \textbf{15.00} & 232.57 &\textbf{ 15.00} & 232.57 & \textbf{15.00}&\textbf{159.19} &-& HS & BRL\\ 
  Highly Dense (XXIII) & \textbf{640.62} & \textbf{15.00} & 646.74 & \textbf{15.00} & 646.74 & \textbf{15.00} & 644.63 & \textbf{15.00} & 12062.78 & - & BRL & BRL \\ 
  Highly Dense (XXIV) & 20.43 & 26.00 & \textbf{18.65} & 47.00 & 18.96 & 47.00 & 21.62 & \textbf{50.00} & 193.81 & - & BRL & rLASSO\\ 
\bottomrule
\end{tabular}}
\end{center}
\end{table}

\subsection{Real Data Applications}
\subsubsection{Parameter Estimation}

To illustrate parameter estimation in real data, we pay a revisit to the prostate cancer dataset \citep{Stamey1989}. This dataset has been analyzed by many previous authors \citep{Tibshirani1996, Zou2005, Park2008, Li2010, Mallick2018, Kyung2010}, and is available from the R package \code{ElemStatLearn}. Briefly, it contains $n=97$ prostate specific antigen (PSA) measurements from prostate cancer patients who were about to receive a radical prostatectomy. PSA is a protein that is produced by the prostate gland, with a higher level indicating a greater chance of having prostate cancer. The goal is to predict the log of PSA (\emph{lpsa}) from a number of clinical measurements ($p=8$) including (i) log cancer volume (\emph{lcavol}), (ii) log prostate weight (\emph{lweight}), (iii) clinical age of the patient (\emph{age}), (iv) log of benign prostatic hyperplasia amount (\emph{lbph}), (v) seminal vesicle invasion (\emph{svi}), (vi) log of capsular penetration (\emph{lcp}), (vii) Gleason score (\emph{gleason}), and (viii) percent of Gleason scores 4 or 5 (\emph{pgg45}).\par

\begin{figure}
\begin{center}
\includegraphics[width=0.9\textwidth]{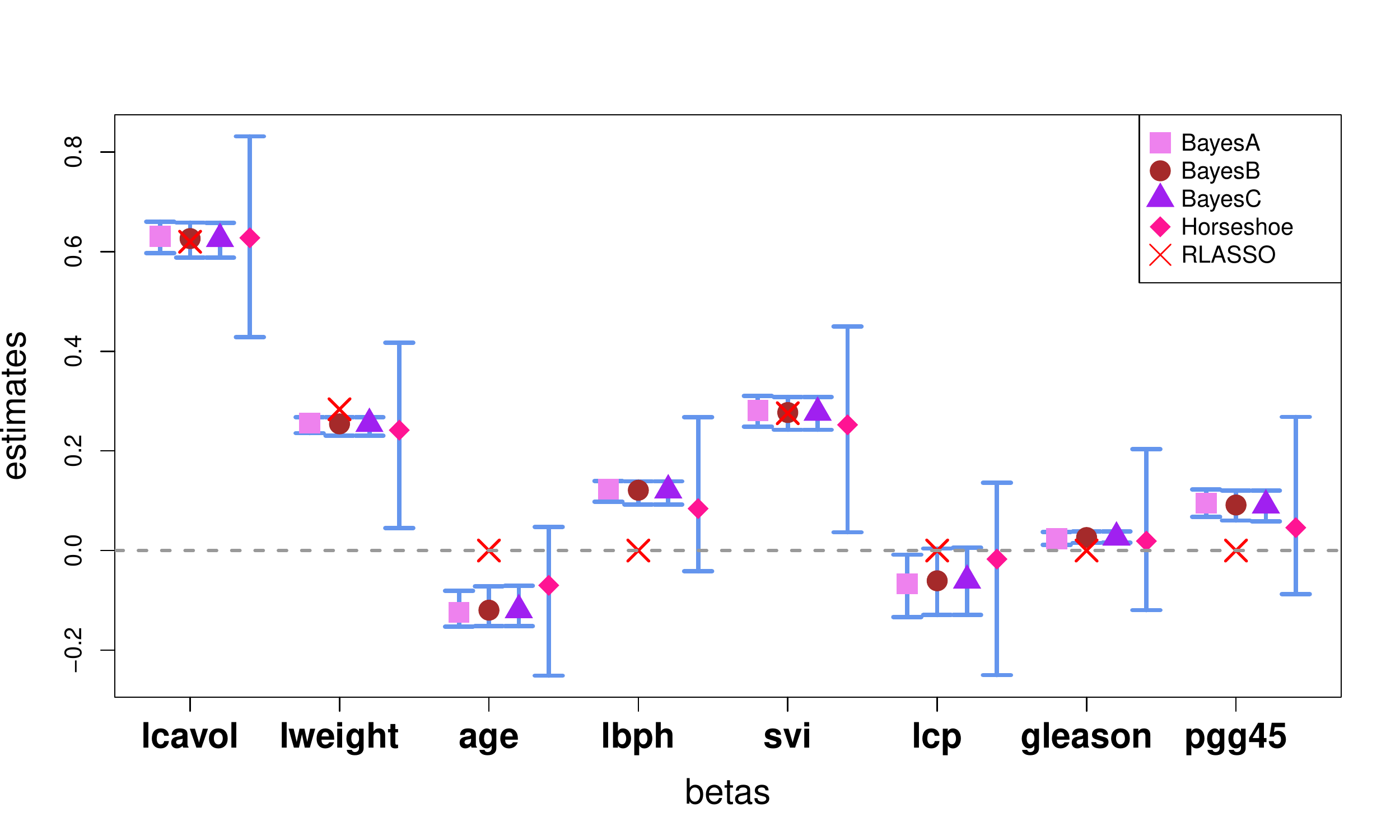}
\caption{Posterior mean and 95$\%$ credible intervals for the 8 covariates in the Prostate cancer dataset with overlaid frequentist rLASSO estimates.}\label{figure2}
\end{center}
\end{figure}

We first compare the classical and Bayesian rLASSO estimates using the full dataset. The response is centered and the predictors are standardized as a standard preprocessing step prior to modeling. We find that the classical rLASSO solution does not always coincide with the joint mode of the posterior distribution (Figure \ref{figure2}). This can be explained by the fact that the full Bayes estimator solves a fundamentally different objective function by marginalizing over the hyperparameters, leading to a considerably different estimate than the classical solution, paralleling the findings of \citet{Park2008}, \citet{Hans2009}, and \citet{Polson2014}. \par

Specifically, with the exception of $age$, $lbph$, and $pgg45$ (all aggressively zeroed out by the classical method despite showing evidence of appreciable posterior uncertainty), the final Bayesian and classical model fits are nearly identical, indicating that the Bayesian rLASSOs, in general, mimic the frequentist method well, with the added benefit of automatic standard error estimation, as an effortless byproduct of the corresponding MCMC procedures. As it is important to account for model uncertainty in prediction while achieving model selection, the posterior mean estimator under the rLASSO prior is particularly appealing.\par 

To investigate the possible susceptibility of the results with respect to $\lambda$, we also conducted additional numerical experiments by considering several choices of the hyperprior which covers a range of non-informative and weakly informative settings (Table \ref{table3}). This sensitivity analysis reveals that the coefficient estimates of the Bayesian rLASSO remains more or less robust across hyperparameter tuning (Table \ref{table3}), which is consistent with the reasoning that parameters that are deeper in the hierarchy have less effect on the inference \citep{Leng2014}. Further comparison with the horseshoe prior revealed that the Bayesian rLASSO estimator usually provides narrower credible intervals and promotes less aggressive shrinkage (Figure \ref{figure2}). All the three methods favorably conclude that the variables \emph{lcavol}, \emph{lweight}, and \emph{svi} are the most important predictors of PSA, meaning that the substantive conclusion based on the `top hits' will be quite similar no matter which approach is used.


\begin{table}[ht]
\begin{center}
\caption{Posterior mean coefficient estimates of the Prostate cancer covariates across a range of hyperpriors for the reciprocal Bayesian LASSO. Gamma$(2, 2)$ was considered by \citet{Polson2014} for the Bayesian bridge estimator, whereas, \citet{Kyung2010} considered the Gamma$(1, 0.01)$ hyperprior for a variety of Bayesian shrinkage priors. In addition to the default Gamma$(0.001, 0.001)$ hyperprior, we also investigated the effect of three weakly informative priors on $\lambda$ (Gamma$(5, 5)$, Gamma$(10, 10)$, and Gamma$(20, 20)$), which confirmed that the reciprocal Bayesian LASSO estimates are not sensitive to the hyperparameter choices.} \label{table3}
\begin{tabular}{rrrrrrrrr}
  \hline
 \textbf{$\lambda \sim \text{Gamma}(a,b)$} & \textbf{lcavol} & \textbf{lweight} & \textbf{age} & \textbf{lbph} & \textbf{svi} & \textbf{lcp} & \textbf{gleason} & \textbf{pgg45} \\ 
  \hline
  $a = 0.001$, $b = 0.001$ & 0.63 & 0.25 & -0.12 & 0.12 & 0.28 & -0.06 & 0.03 & 0.09 \\ 
  $a = 2$, $b = 2$ & 0.63 & 0.25 & -0.12 & 0.12 & 0.28 & -0.06 & 0.03 & 0.09 \\ 
  $a = 1$, $b = 0.01$ & 0.63 & 0.25 & -0.12 & 0.12 & 0.28 & -0.06 & 0.03 & 0.09 \\ 
  $a = 5$, $b = 5$ & 0.64 & 0.30 & -0.22 & 0.20 & 0.34 & -0.19 & 0.13 & 0.12 \\ 
  $a = 10$, $b = 10$ & 0.63 & 0.25 & -0.12 & 0.12 & 0.28 & -0.06 & 0.03 & 0.09 \\ 
  $a = 20$, $b = 20$ & 0.63 & 0.25 & -0.12 & 0.12 & 0.28 & -0.06 & 0.02 & 0.09 \\ 
   \hline
\end{tabular}
\end{center}
\end{table}

\subsubsection{Out-of-sample Prediction Accuracy}

For the purpose of comparison with frequentist rLASSO based on out-of-sample prediction accuracy, we next validate our method across a series of publicly available datasets from diverse biomedical application domains. In particular, in addition to the Prostate Cancer (PC) data, we analyze three more datasets representative of both low-dimensional and high-dimensional regression settings. We employ a repeated 5-fold cross-validation procedure, in which, we randomly partition the data into five equal folds, iteratively taking each fold as test set and the rest as training set and compute the median MSE as our mean squared prediction error (MSPE), wherein models are fitted on the training set and the mean squared error of the residuals are calculated on the test set. We apply the same set of competing methods and the same choice of hyperparameters as before.\par 

The first high-dimensional data we analyze is the obesity microbiome dataset \citep{Goodrich2014}, available from \citet{Duvallet2017}, containing fecal samples from $n = 414$ individuals from the TwinsUK population, including 135 obese cases and 279 controls. Raw sequencing data for this study was processed through a standardized pipeline, as described in \citet{Duvallet2017}, yielding a total of $11, 225$ microbial taxonomic features (OTUs). Following the quality control recommendations provided in \citet{Duvallet2017}, we discard OTUs present in $<5\%$ of the samples. OTUs are further collapsed to the genus level by summing their respective relative abundances, discarding any OTUs that are unannotated, resulting in a smaller number of highly informative features ($p = 99$, including individual OTUs and aggregated elements of the taxonomy) for final modeling \citep{Zhou2019}. Our goal is to predict body mass index (BMI) directly as a continuous phenotype for the same 414 individuals using the relative abundances of $p = 99$ derived microbial features.\par

In addition to the obesity microbiome dataset (GR) described above, we also assess predictive performance in a high-dimensional gene expression data related to the gene TGFB, which encodes a secreted ligand of the Transforming growth factor beta (TGFB) superfamily of proteins that control proliferation, differentiation, and other functions in many cell types \citep{Calon2012}. Recently, \citet{Calon2012} used mice experiments to identify $p = 172$ TGFB-related genes potentially related to colon cancer progression and further validated these genes in an independent cohort of $n = 262$ human patients. The response variable of interest is the overall TGFB level (average log-transformed expression of TGFB1, TGFB2 and TGFB3 mRNAs) in a given sample (a surrogate for colon cancer progression), which we want to predict based on the gene expression of other TGFB-related genes ($p = 172$). We refer to this dataset as CC, which is available from \citet{Rossell2017}. We additionally consider the benchmark Diabetes (DB) dataset \citep{Efron2004} available from the R package \code{lars}, thus showcasing the applicability of our method across both small and large datasets.\par

The cross-validation results are summarized in Table \ref{table4}. The conclusions are similar to those reported in Table \ref{table2}. In most cases, the reciprocal Bayesian LASSO has better predictive performance, outperforming both competing methods in estimation and prediction, while performing on par with respect to parsimonious model selection. Overall, these findings suggest that the reciprocal Bayesian LASSO together with the frequentist backpropagation is extremely effective in detecting a parsimonious subset of predictors while maintaining a low out-of-sample prediction error, improving upon the performance of frequentist rLASSO.


\begin{table}
\begin{center}
\caption{Mean squared prediction error (MSPE) and model size (MS) based on 5-fold CV for the prostate cancer (PC), diabetes (DB), obesity microbiome (GR), and Colon Cancer (CC) datasets.  Predictions are based on the posterior mean for the Bayesian methods and on the model fitting by the S5 algorithm by \citet{Shin2018} for the frequentist rLASSO. For each dataset, the best method ($\text{Winner}$) is also reported. The bold numbers are the minimum out-of-sample prediction accuracy (MSPE) or minimum non-zero model size (MS) per scenario. BRL stands for Bayesian Reciprocal LASSO, whereas, HS is an abbreviation for the horseshoe method. For simplicity, we declare BRL as the winner in the event of a tie involving any BRL method.} \label{table4}
\scalebox{0.6}{
        \begin{tabular}{lrrrrrrrrrrrrrr}\toprule
          &\multicolumn{2}{c}{\textbf{Dimension}}&\multicolumn{2}{c}{\textbf{BayesA}}&\multicolumn{2}{c}{\textbf{BayesB}}&\multicolumn{2}{c}{\textbf{BayesC}}&\multicolumn{2}{c}{\textbf{rLASSO}}&\multicolumn{2}{c}{\textbf{Horseshoe (HS) }}&\multicolumn{2}{c}{\textbf{Winner}}
            \\\cmidrule(r){2-3}\cmidrule(r){4-5}\cmidrule(r){6-7}\cmidrule(r){8-9}\cmidrule(r){10-11}  \cmidrule(r){12-13} \cmidrule(r){14-15}
              \textbf{Dataset} & $n$ & $p$ & MSPE & MS & MSPE & MS & MSPE & MS & MSPE & MS & MSPE & MS & MSPE & MS\\\midrule
PC & 97 & 8 & 0.5913 &	5.6&	0.5520&	1.8&	0.5520&	1.8&	\textbf{0.5063}&	2.8&	0.6631&	\textbf{1.0} & rLASSO & HS\\
DB & 442 & 10 & \textbf{2831.74}&	\textbf{10.0}&	\textbf{2831.74}&	\textbf{10.0}&	\textbf{2831.74}&	\textbf{10.0}&	\textbf{2831.74}&	\textbf{10.0}&	2833.71&	\textbf{10.0} & BRL & BRL\\
GR & 414 & 99 & 42.7293 &	30.0&	\textbf{40.6481}&	\textbf{29.6}&\textbf{40.6481}&	\textbf{29.6}&	40.9289&	30.0&	57.7048&	66.2 & BRL & BRL\\
CC & 262 & 172 & \textbf{0.5040}&	10.6&	0.6979&	12.4&	0.5375&	12.4&	0.5047&	\textbf{4.0}&	0.9965&	0.0 & BRL & rLASSO\\
\bottomrule
\end{tabular}}
\end{center}
\end{table}


\section{Extensions} \label{section6}
The hierarchies of Section \ref{section4} can be used to mimic or implement many other regularization methods through a carefully-specified prior on $\beta$. We briefly describe Bayesian equivalents of some as-yet-unproposed reciprocal methods. Additional extensions are presented in the Appendix.

\subsection{Reciprocal bridge regularization}
One immediately obvious extension of the reciprocal LASSO is the `reciprocalized' version of the bridge regression \citep{Frank1993}, which solves the following problem:
$$
\underset{\bm{{\beta}} }{\text{arg min}} \ (\bm{y}-X\bm{\beta})'(\bm{y}-X\bm{\beta}) + \lambda\sum_{j=1}^{p}\frac{1}{|{\beta}_{j}|^{\alpha}} I\{\beta_j \neq 0\},
$$
for some $\alpha \geq 0$ ($\alpha = 0$ corresponds to an $L_0$ penalty), reducing to rLASSO when $\alpha = 1$. The Bayesian analogue of this penalization involves using a prior on $\beta$ of the form
\begin{equation} \label{himel5}
\pi(\beta) = \frac{{\lambda}^{\frac{1}{\alpha}}}{2\beta^2\Gamma{(\frac{1}{\alpha}+1)}}\exp\{-\frac{\lambda}{|\beta|^{\alpha}}\}I\{\beta\neq 0\},
\end{equation}
where $\alpha>0$ is a shape parameter and $\lambda>0$ is a scale parameter. We refer to this distribution as Inverse Generalized Gaussian (IGG) distribution. It is to be noted that when $\alpha = 2$, it reduces to a reciprocal ridge prior induced by independent inverse normal distributions on the coefficients. \citet{Robert1991} investigated the properties of this class of priors as a conjugate prior family in a normal estimation problem. Using a similar decomposition as Proposition 1, we have the following:
$$
\frac{{\lambda}^{\frac{1}{\alpha}}}{2\beta^2\Gamma{(\frac{1}{\alpha}+1)}}e^{-\lambda|\beta|^{-\alpha}}= \frac{{\lambda}^{\frac{1}{\alpha}}}{2\beta^2\Gamma{(\frac{1}{\alpha}+1)}}\int_{u > |\beta|^{-\alpha}} \lambda e^{-\lambda u} du
$$
$$ 
= \int_{0}^{\infty} \underbrace{\frac{\eta^{\frac{1}{\alpha}}}{2\beta^2} I\{|\beta| > \eta^{\frac{1}{\alpha}}\}}_{\text{Double Pareto} (x_m = \eta^{\frac{1}{\alpha}}, \ \psi = 1)}\underbrace{\frac{\lambda^{(\frac{1}{\alpha}+1)}}{\Gamma(\frac{1}{\alpha}+1)} \eta^{-(\frac{1}{\alpha}+1)-1}\exp(\frac{-\lambda}{\eta})}_{\text{Inverse Gamma} (\frac{1}{\alpha}+1, \ \lambda)} d\eta.
$$
The hierarchical representations of the types in Section \ref{section3} can be achieved by placing appropriate independent distributions on the corresponding latent variables. \par

The penalty function in \eqref{himel1} can also be made `adaptive' by choosing variable-specific tuning parameters as follows:
$$
\underset{\bm{{\beta}} }{\text{arg min}} \ (\bm{y}-X\bm{\beta})'(\bm{y}-X\bm{\beta}) + \sum_{j=1}^{p}\frac{\lambda_{j}}{|{\beta}_{j}|^{\alpha}} I\{\beta_j \neq 0\},
$$
where $\lambda_j>0$ is the tuning parameter for the $j^\text{th}$ coefficient, which can be effortlessly appended (for estimation purposes) in the corresponding MCMC algorithm without significantly increased computational burden, unlike the frequentist framework which must solve a multi-hyperparameter optimization problem to estimate $\bm{\beta}$.

\subsection{Extensions to logistic and quantile regression}
The proposed methods can be extended to logistic regression (LR) and quantile regression (QR) by adding another layer of hierarchy in the corresponding MCMC algorithms. Briefly, this can be achieved by introducing a second set of latent variables to represent the corresponding LR and QR likelihoods as mean variance mixtures of Gaussian models with respect to known mixing measures, while attempting to minimize the following penalized likelihoods:
$$
{\hat{\bm{\beta}}}_{\text{LR}} = \underset{\bm{\beta}}{\text{arg min}} \sum_{i=1}^{n} \log(1 + \exp \{-y_i x_i^{'} \bm{\beta}\}) + \ \lambda \sum_{j=1}^{p}\frac{1}{|{\beta}_{j}|} I\{\beta_j \neq 0\}, \ \bm{y} \in \{ \pm 1\},
$$
$$
{\hat{\bm{\beta}}}_{\text{QR}} = \underset{\bm{\beta}}{\text{arg min}} \sum_{i=1}^{n} \{|y_i - x_i^{'} \bm{\beta}| + (2q - 1) (y_i - x_i^{'} \bm{\beta})\} + \ \lambda \sum_{j=1}^{p}\frac{1}{|{\beta}_{j}|} I\{\beta_j \neq 0\}, \ q \in (0,1).
$$ 
We skip the Bayes construction details due to space constraints, and refer the readers to the corresponding distributional theory presented in \citet{Polson2013a} and \citet{Polson2013b}.
\subsection{Reciprocal Bayesian LASSO for general models}
Finally, within the realm of reasonable large-sample approximation, similar MCMC algorithms can be used to fit Bayesian analogues of rLASSO-penalized general regression models, extending the reciprocal Bayesian LASSO to more complex models such as generalized linear models (GLMs), Cox’s models, non-zero-inflated and zero-inflated count models, and so on. Let us denote by $L(\bm{\beta})$ the negative log-likelihood. Following \citet{Wang2007}, $L(\bm{\beta})$ can be approximated by least squares approximation (LSA) as follows:
$$
L(\bm{\beta})  \approx \frac{1}{2}(\bm{\beta}- \tilde{\bm{\beta}})'\hat {\Sigma}(\bm{\beta}- \tilde{\bm{\beta}}),
$$
where $\tilde{\bm{\beta}}$ is the MLE of $\bm{\beta}$ and ${\hat {\Sigma}}^{-1}=\delta^2 L(\bm{\beta})/\delta \bm{\beta}^2$. Therefore, for a general model, the conditional distribution of $\bm{y}$ is given by
$$
\bm{y}|\bm{\beta} \sim \text{exp} \left\{-\frac{1}{2}(\bm{\beta}- \tilde{\bm{\beta}})'\hat {\Sigma}(\bm{\beta}- \tilde{\bm{\beta}})\right \}.
$$
This allows the general likelihoods to be similarly represented using the hierarchies introduced in Section \ref{section3}, yielding tractable full conditional distributions. As a cautionary note, we consider this one-step LSA-based extension as only an approximate solution for general models as better solutions exist to enable exact MCMC computation for non-Gaussian likelihoods, for example, using the dynamic Hamiltonian Monte Carlo algorithm implemented in the Stan software \citep{Gelman2015}, as described in \citet{Piironen2017} and \citet{Gelman2003}.
 

\section{Conclusions} \label{section7}
We have described a series of Bayesian methods that allow practitioners to estimate the full joint distribution of regression coefficients under the reciprocal LASSO model. Our formulation obtained through a particular scale mixture of inverse uniform densities combines the best of both worlds in that (i) fully Bayes inference is feasible through its hierarchical representation, providing a measure of uncertainty in estimation, (ii) while the implementation of a posthoc sparsification method distills the potentially high-dimensional Bayesian posterior distribution into a simple, interpretable model. Given the excellent performance in a variety of simulation studies and real data applications, the reciprocal Bayesian LASSO should be useful as a nonlocal prior in a broad variety of settings.  \par

On the practical side, these methods can be implemented in commercial software with minimal programming effort. They can also be readily extended to several other penalties, providing a unified  framework for reciprocal regularization. Being directly based on Bayesian hierarchical formulation, our approach has two major advantages: (i) its conceptual simplicity within a well-established framework and (ii) its transportability to other models (e.g., extensions to binary and quantile reciprocal LASSO regression for which no frequentist solutions exist). This significantly expands the scope of reciprocal LASSO by setting the ground for future methodological advances. To facilitate reproducibility and replication, an R package implementing these methods is made publicly available on the first author's GitHub website. \par

We envision several computational and statistical fine-tunings that may further improve the practical applicability of our approach. Going beyond the scope of this work, we anticipate data-driven specification of the global shrinkage parameter ($\lambda$) based on prior domain knowledge, inspired by similar methodological developments in the Bayesian regularization literature \citep{Piironen2017}. This will enable practitioners to control the effective number of non-zero parameters through the rLASSO prior, while also allowing a modest amount of regularization for the largest signals. The hierarchical global-local scale mixture construction proposed in this work can also be encapsulated in an elliptical slice sampler framework to produce a substantially better effective sample size per second \citep{Hahn2019}. Combined, such extensions will allow researchers to facilitate scalable adaptation of the Bayesian reciprocal LASSO in diverse applications, moving beyond routine analysis towards real-time uncertainty quantification \citep{Johndrow2020}.\par

While the technical groundwork in this paper focuses on Bayesian variable selection without a spike-and-slab \citep{Shi2019, Miao2020} or model averaging \citep{Leng2014} component, both these paradigms represent promising next steps for future research. Alternative to the MCMC-based approaches discussed here, one may also consider variational algorithms, which can significantly reduce the computational bottlenecks associated with the fully Bayesian approaches, although at the cost of being less accurate and more sensitive to parameter initialization \citep{Bai2018}. A computationally efficient adaptation of our approach to GLMs and survival models (i.e. without enforcing local approximation strategies such as the LSA) may yield further advantages. Similar to the resurgence of Bayesian LASSO-inspired methodological enhancements in the last decade or so \citep{Van-Erp2019, Mallick2013}, our framework opens up avenues for further research in developing theoretical insights as well as computational advancements in many more interesting problems such as shrinkage of basis coefficients in nonparametric regression and covariance matrix estimation and settings such as multivariate longitudinal analysis, factor analysis, and nonparametric Bayes modeling, among others. We thus hope to see a rapid expansion in both the scalability and applicability of reciprocal Bayesian LASSO in future studies.

\section*{Code Availability}
The implementation of \code{BayesRecipe} is publicly available with source code, documentation, tutorial data, and as an R package at \url{https://github.com/himelmallick/BayesRecipe}. R codes to generate figures and results from this manuscript are available from the first author upon request.

\section*{Data Availability}
The prostate cancer dataset is publicly available from the R package \code{ElemStatLearn}. The diabetes dataset is available from the R package \code{lars}. The remaining datasets are available from the R package \code{BayesRecipe} at \url{https://github.com/himelmallick/BayesRecipe}.

\section*{Acknowledgements}
We sincerely thank Dr. Minsuk Shin (University of South Carolina) for providing the skeleton code for solving the frequentist reciprocal LASSO using a modified version of the S5 algorithm. We would also like to thank Drs. Devan Mehrotra (Merck), Yue-Ming Chen (Merck) and Jeong Hwan Kook (Merck) for offering constructive feedback on an earlier version of the manuscript. The authors would like to thank the editor and two anonymous referees who kindly provided valuable comments and suggestions that greatly improved the quality of the article. Finally, we would like to acknowledge Dr. Ray Bai (University of South Carolina) for serving as an arXiv endorser for the preprint version of the paper (arXiv:2001.08327).

\clearpage
\section*{Appendix}
\beginsupplement

\section*{A. Proofs}
\subsection*{A.1. SMDP Representation}

Let $\beta \sim \text{DP}(x_m, \psi)$ denote that $\beta$ has a symmetric double Pareto (or inverse Uniform) density satisfying the following definition.\ \\

\textbf{Definition 1} \emph{A random variable $\beta$ with scale parameter $x_m>0$ and shape parameter $\psi>0$ follows a double Pareto (type I) distribution if its density function $f$ is of the form  $f(\beta)= \frac{\psi x_m^{\psi}}{2\beta^{(\psi + 1)}} I\{|\beta| \geq x_m\}$, where $I(.)$ denotes an indicator function.}\ \\

\textbf{Proof of Proposition 1:} For an inverse Laplace distribution with scale parameter $\lambda >0$, the following is obvious
$$
\frac{\lambda}{2\beta^2}e^{-\lambda|\beta|^{-1}}= \frac{\lambda}{2\beta^2} \int_{u > |\beta|^{-1}} \lambda e^{-\lambda u} du.
$$
Consider the transformation, $u \to \eta \equiv u^{-1}$, which implicitly absorbs a factor of $\eta^{-1}$ from the normalization constant of the double Pareto kernel into the inverse gamma conditional for $\eta$ as follows:
$$ 
=\int_{0}^{\infty} \underbrace{\frac{\eta}{2\beta^2}I\{|\beta| > \eta\}}_{\text{Double Pareto} (x_m = \eta, \ \psi = 1)}\underbrace{\frac{\lambda^2}{\Gamma(2)} \eta^{-2-1}\exp(\frac{-\lambda}{\eta})}_{\text{Inverse Gamma} (2, \ \lambda)} d\eta.
$$
This proves Proposition 1. 

\subsection*{A.2. SMTN Representation} 
In order to derive the SMTN representation of the inverse Laplace density, we introduce the following definition and lemma.\ \\ 

\textbf{Definition 2} \emph{A random variable $\beta$ with location parameter $\mu$, scale parameter $\xi$, and shape parameter $\alpha$ follows a generalized double Pareto distribution if its density function $f$ is of the form  $f(\beta)= \frac{1}{2\xi} \left (1 + \frac{|\beta| - \mu}{\alpha\xi}\right)^{-(\alpha+1)}$, where $|\beta| \geq \mu$, $\alpha>0$, $\xi >0$, and $\mu \in \mathbb{R}$.}\ \\

Let $\beta \sim \text{GDP}(\mu, \xi, \alpha)$ denote that $\beta$ has a generalized double Pareto density with location parameter $\mu$, scale parameter $\xi$, and shape parameter $\alpha$, where $|\beta| \geq \mu$, $\alpha>0$, $\xi >0$, and $\mu \in \mathbb{R}$. \citet{Armagan2013a} derived the SMN representation of the GDP distribution for $\mu = 0$. Here we extend the result for general $\mu$ and provide the following lemma.

\begin{lemma}
\emph{Let $\beta \sim N(0, \zeta)I(|\beta| > \mu)$, $\zeta \sim Exp(\lambda^2/2)$, and $\lambda \sim Gamma (\alpha, \eta)$, where $\alpha>0$ and $\eta>0$. The resulting marginal density for $\beta$ is $\text{GDP}(\mu, \xi = \eta/\alpha, \alpha).$}\
\end{lemma}

\textbf{Proof of Proposition 2}: By virtue of Lemma 1, the proof is an immediate consequence of the fact that a $\text{GDP}(\mu, \xi, \alpha)$ distribution reduces to a double Pareto (type I) distribution $\text{DP}(x_m, \psi)$  with $x_m = \xi/\alpha$, $\psi = 1/\alpha$, and $\mu = \xi/\alpha$. Setting $\alpha=1$ completes the proof.


\section*{B. MCMC Algorithms for the Reciprocal Bayesian LASSO}
\begin{figure}[h]
\begin{center}
\fbox{\parbox{\linewidth}{
\begin{algorithmic}
\Input $(\bm{y}, X)$
\Initialize $(\bm{\beta}, \sigma^2, \bm{u}, \lambda)$
\For{$t=1,\ldots, (t_{\text{max}}+t_{\text{burn-in}})$}
\begin{enumerate}
\item Sample $\bm{u}|. \sim \displaystyle \prod_{j=1}^{p} \text{Exponential}(\lambda) I\{u_j>\frac{1}{|\beta_{j}|}\}$.
\item $\sigma^2|. \sim \text{Inverse Gamma} \ (\frac{n-1}{2}, \frac{1}{2}(\bm{y}-X\bm{\beta})'(\bm{y}-X\bm{\beta}))$.
\item Generate $\bm{\beta}|.$ from a truncated multivariate normal proportional to $$N_p(\hat {\bm{\beta}}_{\text{MLE}},\sigma^2 {(X'X)}^{-1} ) \displaystyle \prod_{j=1}^{p} I\{|\beta_{j}| > \frac{1}{u_j}\}.$$
\end{enumerate}
\EndFor
\Update Hyperparameter $\lambda$ as required
\end{algorithmic}
}}
\caption{The Reciprocal Bayesian LASSO Gibbs Sampler Using SMDP.} \label{figureS1}
\end{center}
\end{figure}


\clearpage
\begin{figure}[h]
\fbox{\parbox{\linewidth}{
\begin{algorithmic}
\Input $(\bm{y}, X)$
\Initialize $(\bm{\beta}, \sigma^2, \bm{\zeta}, \bm{\lambda}, \bm{u}, \lambda)$
\For{$t=1,\ldots, (t_{\text{max}}+t_{\text{burn-in}})$}
\begin{enumerate}
\item Sample $\bm{u}|. \sim \displaystyle \prod_{j=1}^{p} \text{Exponential}(\lambda) I\{u_j>\frac{\sigma}{|\beta_{j}|}\}$.
\item $\bm{\zeta}|. \sim  \displaystyle \prod_{j=1}^{p} \text{Gamma}(2, \left (\frac{|\beta_j|}{\sigma} + \frac{1}{u_j} \right ))$.
\item $\bm{\tau^{-1}}|. \sim \displaystyle \prod_{j=1}^{p} \text{Inverse-Gaussian}(\sqrt{\frac{\zeta_j^{2}\sigma^2}{\beta_j^{2}}}, \zeta_j^{2})$.
\item Generate $\sigma^2|.$ from a truncated inverse gamma proportional to $$\text{Inverse-Gamma}(\frac{n-1+p}{2}, \frac{R + \bm{\beta}^{'}T^{-1}\bm{\beta}}{2}) I\{\sigma^2 < \text{Min}_j(\beta_j^2 u_j^2)\}$$
where $R = (\bm{y}-X\bm{\beta})'(\bm{y}-X\bm{\beta})$ and $T = \text{diag} (\tau_1, \dots, \tau_p)$.
\item Generate $\bm{\beta}|.$ from a truncated multivariate normal proportional to $$N_p((X'X + T^{-1})^{-1} X^{'}\bm{y},\sigma^2 {(X'X + T^{-1})}^{-1}) \displaystyle \prod_{j=1}^{p} I\{|\beta_{j}| > \frac{\sigma}{u_j}\}.$$
\end{enumerate}
\EndFor
\Update Hyperparameter $\lambda$ as required
\end{algorithmic}
}}
\caption{The Reciprocal Bayesian LASSO Gibbs Sampler Using SMTN.} \label{figureS2}
\end{figure}

\section*{C. Additional Extensions}

\subsection*{C.1. An Alternative Formulation Using Reciprocal $n$-monotone Densities}
An alternative representation motivated by the Bayesian bridge formulation of \citet{Polson2014} can also be reinforced for the Bayesian reciprocal bridge prior (described in Section 6). In particular, building on a classic theorem, \citet{Polson2014} showed that any $n$-monotone density $f(x)$ can be represented as a scale mixture of betas and the mixing distribution can be explicitly determined by using the derivatives of $f$. Here we extend the result of Theorem 2.1 of \citet{Polson2014} to reciprocal $n$-monotone densities and introduce the following lemma.

\begin{lemma} 
Let $f(x)$ be a bounded density function that is symmetric about zero and $n$-monotone over $(0, \infty)$, normalized so that $f(0) = 1$. Let $C = \{2\int_{0}^{\infty}f(t)dt\}^{-1}$ denote the normalizing constant that makes $f(x)$ a proper density on the real line. Then the reciprocal of $f$ can be represented as the following mixture for any integer $k$, $1 \leq k \leq n$:
$$
\frac{Cf(\frac{1}{x})}{x^2} = \int_{0}^{\infty} \frac{1}{sx^2} \ k \ {\left\{ 1 -  \frac{1}{s|x|}\right\}}_{+}^{k-1} \ g(s) ds,
$$
where $a_{+} = max (a, 0)$, and where the mixing density $g(s)$ is 
$$
g(s) = Ck^{-1} \sum_{j=0}^{k-1} \frac{(-1)^{j}}{j!} \left\{js^{j} f^{(j)} (s)  + s^{j+1} f^{(j+1)} (s) \right\}.
$$
\end{lemma}

Following \citet{Polson2014}, we refer to the resulting kernel functions as reciprocal (or inverse) Bartlett–Fejer kernels. Using $k=2$ and extending Corollary 1 of \citet{Polson2014}, we have the following result.
\begin{corollary} 
Let $f(x)$ be a function that is symmetric about the origin; integrable, convex, and twice-differentiable on $(0, \infty)$; and for which $f(0) = 1$. Let $C = \{2\int_{0}^{\infty}f(t)dt\}^{-1}$ denote the normalizing constant that makes $f(x)$ a density on the real line. Then the reciprocal of $f$ has the following mixture representation:
$$
\frac{Cf(\frac{1}{x})}{x^2}  = \int_{0}^{\infty} \frac{1}{st^2} \ {\left\{ 1 -  \frac{1}{s|t|}\right\}}_{+} \ Cs^{2}f^{''}(s)ds,
$$
where $a_{+} = max (a, 0)$.
\end{corollary}

Using the corollary, we have the following alternative representation for the reciprocal bridge prior:
$$
\frac{{\lambda}^{\frac{1}{\alpha}}}{2\beta^2\Gamma{(\frac{1}{\alpha}+1)}}e^{-\lambda|\beta|^{-\alpha}}= \int_{0}^{\infty} \frac{\lambda^{\frac{1}{\alpha}}}{\beta^2w^{\frac{1}{\alpha}}} {\left\{ 1 -  \abs*{\frac{\lambda^{\frac{1}{\alpha}}}{\beta w^{\frac{1}{\alpha}}}}\right\}}_{+} p(w| \alpha) dw,
$$
$$
p(w| \alpha) = \frac{1+\alpha}{2} c_{1} w^{1+ \frac{1}{\alpha}} e^{-w} +  \frac{1- \alpha}{2} c_{2} w^{\frac{1}{\alpha}} e^{-w},
$$
where $c_1$ and $c_2$ are the normalizing constants for the component-wise densities in the gamma mixture. The reciprocal Bayesian LASSO naturally arises as a special case, for which the second mixture component vanishes. It is not clear whether an efficient Gibbs sampler can be based on this hierarchy, however.

\subsection*{C.2. A Unified Framework for Reciprocal Shrinkage Densities}
In the proof of Corollary 1 of \citet{Polson2014}, $k=1$ recovers the well-known fact that monotone densities can be expressed as scale mixture of uniforms, which leads to the following characterization based on the scale mixture of inverse uniform or double Pareto densities for the reciprocal shrinkage priors.
\begin{corollary} 
Let $f(x)$ be a function that is symmetric about the origin; integrable, convex, and once-differentiable on $(0, \infty)$; and for which $f(0) = 1$. Let $C = \{2\int_{0}^{\infty}f(t)dt\}^{-1}$ denote the normalizing constant that makes $f(x)$ a density on the real line. Then the reciprocal of $f$ has the following mixture representation:
$$
\frac{Cf(\frac{1}{x})}{x^2}  = \int_{0}^{\infty} \frac{1}{st^2} I\{|t| > \frac{1}{s}\} \ Csf^{'}(s)ds,
$$
where $I(.)$ denotes an indicator function.
\end{corollary}

Applying Corollary 2 to $\pi(\beta) (\beta \in \mathbb{R})$ for which $\pi^{\prime}(\beta)$ exists for all $\beta$, we can define a similar double Pareto scale mixture representation for general reciprocal monotone densities (up to normalizing constants):
\begin{equation} \label{himel6}
\pi(\theta) \propto \pi(\theta|t) \times h(t),
\end{equation}
where $\theta = \frac{1}{\beta}$, $\pi(\theta|t)  = \frac{1}{\text{Uniform}(-t, t)}$, and $h(t) = -2t \times \pi^{\prime}(t)$.\ \\ \par

This implies that there may be many interesting cases where the new approach could be useful, especially in `nonlocalization' or `reciprocalization' of a local shrinkage prior that belongs to the class of monotone densities. Although a long discussion here would lead us astray, we briefly characterize various `reciprocalized' shrinkage priors by capitalizing on \eqref{himel6}. \ \\ \par

One natural candidate for shrinkage priors is the Student’s $t$ distribution with $v \ (v > 0)$ degrees of freedom given by $\pi(\beta) \propto (1 + \frac{\beta^2}{\lambda^2})^{-\frac{(v+1)}{2}}$, the reciprocal of which (i.e. $\pi(\theta), \ \theta = \frac{1}{\beta}$) can be written as follows:
$$
\pi(\theta) \propto \pi(\theta|t) \times h(t),
$$
where $\pi(\theta|t) \propto \frac{1}{\text{Uniform}(-t, t)}$ and $h(t) \propto t^2 (1 + \frac{t^2}{\lambda^2})^{-\frac{(v+3)}{2}}$.\ \\ \par

As the second example, consider the generalized Double Pareto distribution \citep{Armagan2013a} given by $\pi(\beta) \propto (1 + \frac{|\beta|}{\tau})^{-(1+\alpha)}$, which can be reciprocalized as follows
$$
\pi(\theta) \propto \pi(\theta|t) \times h(t),
$$
where $\pi(\theta|t) \propto \frac{1}{\text{Uniform}(-t, t)}$ and $h(t) \propto t (1 + \frac{t}{\tau})^{-(2 + \alpha)}$.\ \\ \par

Finally, consider the horseshoe prior \citep{Carvalho2010}, which does not have a closed form density but has two desirable properties for shrinkage estimation: an infinite spike at zero and heavy tails. Recently, \citet{Wang2013} considered a `logarithm' shrinkage prior that has `horseshoe-like' properties with the added advantage of an explicit density function given by 
$$
\pi(\beta) \propto \log (1 + \frac{\tau^2}{\beta^2}).
$$
It is easy to see that the corresponding `reciprocalized horseshoe-like' prior has the following scale mixture representation:
$$
\pi(\theta) \propto \pi(\theta|t) \times h(t),
$$
where $\pi(\theta|t) \propto \frac{1}{\text{Uniform}(-t, t)}$ and $h(t) \propto (1 + \frac{t^2}{\tau^2})^{-1} I\{t > 0\}$.


\clearpage
\bibliographystyle{plainnat}
\bibliography{BayesRLasso}

\begin{thebibliography}{55}
\providecommand{\natexlab}[1]{#1}
\providecommand{\url}[1]{\texttt{#1}}
\expandafter\ifx\csname urlstyle\endcsname\relax
  \providecommand{\doi}[1]{doi: #1}\else
  \providecommand{\doi}{doi: \begingroup \urlstyle{rm}\Url}\fi

\bibitem[Armagan et~al.(2013)Armagan, Dunson, and Lee]{Armagan2013a}
Artin Armagan, David~B Dunson, and Jaeyong Lee.
\newblock Generalized double {P}areto shrinkage.
\newblock \emph{Statistica Sinica}, 23\penalty0 (1):\penalty0 119, 2013.

\bibitem[Bai and Ghosh(2018)]{Bai2018}
Ray Bai and Malay Ghosh.
\newblock On the beta prime prior for scale parameters in high-dimensional
  {B}ayesian regression models.
\newblock \emph{arXiv preprint arXiv:1807.06539}, 2018.

\bibitem[Bai et~al.(2020)Bai, Rockova, and George]{Bai2020}
Ray Bai, Veronika Rockova, and Edward~I George.
\newblock Spike-and-slab meets {LASSO}: A review of the spike-and-slab lasso.
\newblock \emph{arXiv preprint arXiv:2010.06451}, 2020.

\bibitem[Bhadra et~al.(2019)Bhadra, Datta, Polson, and Willard]{Bhadra2019}
Anindya Bhadra, Jyotishka Datta, Nicholas~G Polson, and Brandon Willard.
\newblock {LASSO} meets horseshoe: A survey.
\newblock \emph{Statistical Science}, 34\penalty0 (3):\penalty0 405--427, 2019.

\bibitem[Bhattacharya et~al.(2016)Bhattacharya, Chakraborty, and
  Mallick]{Bhattacharya2016}
Anirban Bhattacharya, Antik Chakraborty, and Bani~K Mallick.
\newblock Fast sampling with {G}aussian scale mixture priors in
  high-dimensional regression.
\newblock \emph{Biometrika}, 103\penalty0 (4):\penalty0 985--991, 2016.

\bibitem[Brodersen et~al.(2010)Brodersen, Ong, Stephan, and
  Buhmann]{Brodersen2010}
Kay~Henning Brodersen, Cheng~Soon Ong, Klaas~Enno Stephan, and Joachim~M
  Buhmann.
\newblock The balanced accuracy and its posterior distribution.
\newblock In \emph{20th International Conference on Pattern Recognition}, pages
  3121--3124, 2010.

\bibitem[Calon et~al.(2012)Calon, Espinet, Palomo-Ponce, Tauriello, Iglesias,
  C{\'e}spedes, Sevillano, Nadal, Jung, Zhang, et~al.]{Calon2012}
Alexandre Calon, Elisa Espinet, Sergio Palomo-Ponce, Daniele~VF Tauriello, Mar
  Iglesias, Mar{\'\i}a~Virtudes C{\'e}spedes, Marta Sevillano, Cristina Nadal,
  Peter Jung, Xiang H-F Zhang, et~al.
\newblock Dependency of colorectal cancer on a {TGF}-$\beta$-driven program in
  stromal cells for metastasis initiation.
\newblock \emph{Cancer Cell}, 22\penalty0 (5):\penalty0 571--584, 2012.

\bibitem[Carvalho et~al.(2010)Carvalho, Polson, and Scott]{Carvalho2010}
C.~M. Carvalho, N.~G. Polson, and J.~G. Scott.
\newblock The horseshoe estimator for sparse signals.
\newblock \emph{Biometrika}, 97(2):\penalty0 465--480, 2010.

\bibitem[Duvallet et~al.(2017)Duvallet, Gibbons, Gurry, Irizarry, and
  Alm]{Duvallet2017}
Claire Duvallet, Sean~M Gibbons, Thomas Gurry, Rafael~A Irizarry, and Eric~J
  Alm.
\newblock Meta-analysis of gut microbiome studies identifies disease-specific
  and shared responses.
\newblock \emph{Nature Communications}, 8\penalty0 (1):\penalty0 1784, 2017.

\bibitem[Efron et~al.(2004)Efron, Hastie, Johnstone, and Tibshirani]{Efron2004}
B.~Efron, T.~Hastie, I.~Johnstone, and R.~Tibshirani.
\newblock Least angle regression.
\newblock \emph{Annals of Statistics}, 32(2):\penalty0 407--99, 2004.

\bibitem[Emmert-Streib and Dehmer(2019)]{Emmert-Streib2019}
Frank Emmert-Streib and Matthias Dehmer.
\newblock High-dimensional {LASSO}-based computational regression models:
  regularization, shrinkage, and selection.
\newblock \emph{Machine Learning and Knowledge Extraction}, 1\penalty0
  (1):\penalty0 359--383, 2019.

\bibitem[Frank and Friedman(1993)]{Frank1993}
I.~Frank and J.~H. Friedman.
\newblock A statistical view of some chemometrics regression tools (with
  discussion).
\newblock \emph{Technometrics}, 35\penalty0 (2):\penalty0 109--135, 1993.

\bibitem[Gelman et~al.(2003)Gelman, Carlin, Stern, and Rubin]{Gelman2003}
A.~Gelman, J.~Carlin, H.~Stern, and D.~Rubin.
\newblock \emph{Bayesian {D}ata {A}nalysis}.
\newblock Chapman \& Hall, London, 2003.

\bibitem[Gelman et~al.(2015)Gelman, Lee, and Guo]{Gelman2015}
Andrew Gelman, Daniel Lee, and Jiqiang Guo.
\newblock {S}tan: A probabilistic programming language for {B}ayesian inference
  and optimization.
\newblock \emph{Journal of Educational and Behavioral Statistics}, 40\penalty0
  (5):\penalty0 530--543, 2015.

\bibitem[Goodrich et~al.(2014)Goodrich, Waters, Poole, Sutter, Koren, Blekhman,
  Beaumont, Van~Treuren, Knight, and Bell]{Goodrich2014}
Julia~K Goodrich, Jillian~L Waters, Angela~C Poole, Jessica~L Sutter, Omry
  Koren, Ran Blekhman, Michelle Beaumont, William Van~Treuren, Rob Knight, and
  Jordana~T Bell.
\newblock Human genetics shape the gut microbiome.
\newblock \emph{Cell}, 159\penalty0 (4):\penalty0 789--799, 2014.

\bibitem[Hahn and Carvalho(2015)]{Hahn2015}
P~Richard Hahn and Carlos~M Carvalho.
\newblock Decoupling shrinkage and selection in {B}ayesian linear models: a
  posterior summary perspective.
\newblock \emph{Journal of the American Statistical Association}, 110\penalty0
  (509):\penalty0 435--448, 2015.

\bibitem[Hahn et~al.(2019)Hahn, He, and Lopes]{Hahn2019}
P~Richard Hahn, Jingyu He, and Hedibert~F Lopes.
\newblock Efficient sampling for {G}aussian linear regression with arbitrary
  priors.
\newblock \emph{Journal of Computational and Graphical Statistics}, 28\penalty0
  (1):\penalty0 142--154, 2019.

\bibitem[Hans(2009)]{Hans2009}
C.~M. Hans.
\newblock Bayesian {LASSO} regression.
\newblock \emph{Biometrika}, 96\penalty0 (4):\penalty0 835--845, 2009.

\bibitem[Hastie et~al.(2019)Hastie, Tibshirani, and Wainwright]{Hastie2019}
Trevor Hastie, Robert Tibshirani, and Martin Wainwright.
\newblock \emph{Statistical learning with sparsity: the {LASSO} and
  generalizations and generalizations}.
\newblock Chapman and Hall/CRC, 2019.

\bibitem[Johndrow et~al.(2020)Johndrow, Orenstein, and
  Bhattacharya]{Johndrow2020}
James~E Johndrow, Paulo Orenstein, and Anirban Bhattacharya.
\newblock Scalable approximate {MCMC} algorithms for the horseshoe prior.
\newblock \emph{Journal of Machine Learning Research}, 21\penalty0
  (73):\penalty0 1--61, 2020.

\bibitem[Johnson and Rossell(2010)]{Johnson2010}
Valen~E Johnson and David Rossell.
\newblock On the use of non‐local prior densities in {B}ayesian hypothesis
  tests.
\newblock \emph{Journal of the Royal Statistical Society: Series B (Statistical
  Methodology)}, 72\penalty0 (2):\penalty0 143--170, 2010.

\bibitem[Johnson and Rossell(2012)]{Johnson2012}
Valen~E Johnson and David Rossell.
\newblock Bayesian model selection in high-dimensional settings.
\newblock \emph{Journal of the American Statistical Association}, 107\penalty0
  (498):\penalty0 649--660, 2012.

\bibitem[Kim(2007)]{Kim2007}
Hea-Jung Kim.
\newblock Moments of a class of internally truncated normal distributions.
\newblock \emph{Communications for Statistical Applications and Methods},
  14\penalty0 (3):\penalty0 679--686, 2007.

\bibitem[Kyung et~al.(2010)Kyung, Gill, Ghosh, and Casella]{Kyung2010}
M.~Kyung, J.~Gill, M.~Ghosh, and G.~Casella.
\newblock Penalized regression, standard errors, and {B}ayesian {LASSO}s.
\newblock \emph{Bayesian Analysis}, 5\penalty0 (2):\penalty0 369--412, 2010.

\bibitem[Leng et~al.(2014)Leng, Tran, and Nott]{Leng2014}
Chenlei Leng, Minh-Ngoc Tran, and David Nott.
\newblock Bayesian adaptive {LASSO}.
\newblock \emph{Annals of the Institute of Statistical Mathematics},
  66\penalty0 (2):\penalty0 221--244, 2014.

\bibitem[Li and Lin(2010)]{Li2010}
Q.~Li and N.~Lin.
\newblock The {B}ayesian elastic net.
\newblock \emph{Bayesian Analysis}, 5(1):\penalty0 151--70, 2010.

\bibitem[Mallick and Yi(2013)]{Mallick2013}
Himel Mallick and Nengjun Yi.
\newblock Bayesian methods for high dimensional linear models.
\newblock \emph{Journal of Biometrics and Biostatistics}, 1:\penalty0 005,
  2013.

\bibitem[Mallick and Yi(2014)]{Mallick2014}
Himel Mallick and Nengjun Yi.
\newblock A new {B}ayesian {LASSO}.
\newblock \emph{Statistics and Its Interface}, 7\penalty0 (4):\penalty0
  571--582, 2014.

\bibitem[Mallick and Yi(2018)]{Mallick2018}
Himel Mallick and Nengjun Yi.
\newblock Bayesian bridge regression.
\newblock \emph{Journal of Applied Statistics}, 45\penalty0 (6):\penalty0
  988--1008, 2018.

\bibitem[Miao et~al.(2020)Miao, Kook, Lu, Guindani, and Vannucci]{Miao2020}
Yinsen Miao, Jeong~Hwan Kook, Yadong Lu, Michele Guindani, and Marina Vannucci.
\newblock Chapter 7 - {S}calable {B}ayesian variable selection regression
  models for count data.
\newblock In Yanan Fan, David Nott, Michael~S. Smith, and Jean-Luc
  Dortet-Bernadet, editors, \emph{Flexible Bayesian Regression Modelling},
  pages 187--219. Academic Press, 2020.

\bibitem[Nikooienejad et~al.(2016)Nikooienejad, Wang, and
  Johnson]{Nikooienejad2016}
Amir Nikooienejad, Wenyi Wang, and Valen~E Johnson.
\newblock Bayesian variable selection for binary outcomes in high-dimensional
  genomic studies using non-local priors.
\newblock \emph{Bioinformatics}, 32\penalty0 (9):\penalty0 1338--1345, 2016.

\bibitem[Nikooienejad et~al.(2020)Nikooienejad, Wang, and
  Johnson]{Nikooienejad2020}
Amir Nikooienejad, Wenyi Wang, and Valen~E Johnson.
\newblock Bayesian variable selection for survival data using inverse moment
  priors.
\newblock \emph{The Annals of Applied Statistics}, 14\penalty0 (2):\penalty0
  809, 2020.

\bibitem[Park and Casella(2008)]{Park2008}
T.~Park and G.~Casella.
\newblock The {B}ayesian {LASSO}.
\newblock \emph{Journal of the American Statistical Association}, 103\penalty0
  (482):\penalty0 681--686, 2008.

\bibitem[Piironen and Vehtari(2017)]{Piironen2017}
Juho Piironen and Aki Vehtari.
\newblock Sparsity information and regularization in the horseshoe and other
  shrinkage priors.
\newblock \emph{Electronic Journal of Statistics}, 11\penalty0 (2):\penalty0
  5018--5051, 2017.

\bibitem[Polson and Scott(2013)]{Polson2013a}
Nicholas~G Polson and James~G Scott.
\newblock Data augmentation for non-{G}aussian regression models using
  variance-mean mixtures.
\newblock \emph{Biometrika}, 100\penalty0 (2):\penalty0 459--471, 2013.

\bibitem[Polson et~al.(2013)Polson, Scott, and Windle]{Polson2013b}
Nicholas~G Polson, James~G Scott, and Jesse Windle.
\newblock Bayesian inference for logistic models using {P}{\'o}lya--gamma
  latent variables.
\newblock \emph{Journal of the American statistical Association}, 108\penalty0
  (504):\penalty0 1339--1349, 2013.

\bibitem[Polson et~al.(2014)Polson, Scott, and Windle]{Polson2014}
Nicholas~G Polson, James~G Scott, and Jesse Windle.
\newblock The {B}ayesian bridge.
\newblock \emph{Journal of the Royal Statistical Society, Series B
  (Methodological)}, 76(4):\penalty0 713--733, 2014.

\bibitem[Robert(1991)]{Robert1991}
Christian Robert.
\newblock Generalized inverse normal distributions.
\newblock \emph{Statistics and Probability Letters}, 11\penalty0 (1):\penalty0
  37--41, 1991.

\bibitem[Rossell and Telesca(2017)]{Rossell2017}
David Rossell and Donatello Telesca.
\newblock Nonlocal priors for high-dimensional estimation.
\newblock \emph{Journal of the American Statistical Association}, 112\penalty0
  (517):\penalty0 254--265, 2017.

\bibitem[Ruli et~al.(2016)Ruli, Sartori, and Ventura]{Ruli2016}
Erlis Ruli, Nicola Sartori, and Laura Ventura.
\newblock Improved {L}aplace approximation for marginal likelihoods.
\newblock \emph{Electronic Journal of Statistics}, 10\penalty0 (2):\penalty0
  3986--4009, 2016.

\bibitem[Sanyal et~al.(2019)Sanyal, Lo, Kauppi, Djurovic, Andreassen, Johnson,
  and Chen]{Sanyal2019}
Nilotpal Sanyal, Min-Tzu Lo, Karolina Kauppi, Srdjan Djurovic, Ole~A
  Andreassen, Valen~E Johnson, and Chi-Hua Chen.
\newblock {GWASinlps}: non-local prior based iterative {SNP} selection tool for
  genome-wide association studies.
\newblock \emph{Bioinformatics}, 35\penalty0 (1):\penalty0 1--11, 2019.

\bibitem[Shi et~al.(2019)Shi, Lim, and Maiti]{Shi2019}
Guiling Shi, Chae~Young Lim, and Tapabrata Maiti.
\newblock Model selection using mass-nonlocal prior.
\newblock \emph{Statistics and Probability Letters}, 147:\penalty0 36--44,
  2019.

\bibitem[Shin et~al.(2018)Shin, Bhattacharya, and Johnson]{Shin2018}
Minsuk Shin, Anirban Bhattacharya, and Valen~E Johnson.
\newblock Scalable {B}ayesian variable selection using nonlocal prior densities
  in ultrahigh-dimensional settings.
\newblock \emph{Statistica Sinica}, 28\penalty0 (2):\penalty0 1053, 2018.

\bibitem[Song(2018)]{Song2018}
Qifan Song.
\newblock An overview of reciprocal ${L}_{1}$‐regularization for high
  dimensional regression data.
\newblock \emph{Wiley Interdisciplinary Reviews: Computational Statistics},
  10\penalty0 (1):\penalty0 e1416, 2018.

\bibitem[Song and Liang(2015)]{Song2015}
Qifan Song and Faming Liang.
\newblock High-dimensional variable selection with reciprocal
  ${L}_{1}$-regularization.
\newblock \emph{Journal of the American Statistical Association}, 110\penalty0
  (512):\penalty0 1607--1620, 2015.

\bibitem[Stamey et~al.(1989)Stamey, Kabalin, McNeal, Johnstone, Freiha,
  Redwine, and Yang]{Stamey1989}
Thomas~A Stamey, John~N Kabalin, John~E McNeal, Iain~M Johnstone, Fuad Freiha,
  Elise~A Redwine, and Norman Yang.
\newblock Prostate specific antigen in the diagnosis and treatment of
  adenocarcinoma of the prostate. {II}. {R}adical prostatectomy treated
  patients.
\newblock \emph{The Journal of urology}, 141\penalty0 (5):\penalty0 1076--1083

\bibitem[Tibshirani(1996)]{Tibshirani1996}
R.~Tibshirani.
\newblock Regression shrinkage and selection via the {LASSO}.
\newblock \emph{Journal of the Royal Statistical Society. Series B
  (Methodological)}, 58\penalty0 (1):\penalty0 267--288, 1996.

\bibitem[Tibshirani(2011)]{Tibshirani2011}
Robert Tibshirani.
\newblock Regression shrinkage and selection via the {LASSO}: a retrospective.
\newblock \emph{Journal of the Royal Statistical Society: Series B (Statistical
  Methodology)}, 73\penalty0 (3):\penalty0 273--282, 2011.

\bibitem[Van~Erp et~al.(2019)Van~Erp, Oberski, and Mulder]{Van-Erp2019}
Sara Van~Erp, Daniel~L Oberski, and Joris Mulder.
\newblock Shrinkage priors for {B}ayesian penalized regression.
\newblock \emph{Journal of Mathematical Psychology}, 89:\penalty0 31--50 
  0022--2496, 2019.

\bibitem[Vidaurre et~al.(2013)Vidaurre, Bielza, and Larranaga]{Vidaurre2013}
Diego Vidaurre, Concha Bielza, and Pedro Larranaga.
\newblock A survey of ${L}_1$ regression.
\newblock \emph{International Statistical Review}, 81\penalty0 (3):\penalty0
  361--387, 2013.

\bibitem[Wang and Leng(2007)]{Wang2007}
H.~Wang and C.~Leng.
\newblock Unified {LASSO} estimation by least squares approximation.
\newblock \emph{Journal of the American Statistical Association}, 102\penalty0
  (479):\penalty0 1039--1048, 2007.

\bibitem[Wang and Pillai(2013)]{Wang2013}
Hao Wang and Natesh~S Pillai.
\newblock On a class of shrinkage priors for covariance matrix estimation.
\newblock \emph{Journal of Computational and Graphical Statistics}, 22\penalty0
  (3):\penalty0 689--707, 2013.

\bibitem[Woo(2009)]{Woo2009}
Jung-Soo Woo.
\newblock Notes on a skew-symmetric inverse double {W}eibull distribution.
\newblock \emph{Journal of the Korean Data and Information Science Society},
  20\penalty0 (2):\penalty0 459--465, 2009.

\bibitem[Zhou and Gallins(2019)]{Zhou2019}
Yi-Hui Zhou and Paul Gallins.
\newblock A review and tutorial of machine learning methods for microbiome host
  trait prediction.
\newblock \emph{Frontiers in Genetics}, 10:\penalty0 579, 2019.

\bibitem[Zou and Hastie(2005)]{Zou2005}
H.~Zou and T.~Hastie.
\newblock Regularization and variable selection via the elastic net.
\newblock \emph{Journal of the Royal Statistical Society. Series B
  (Methodological)}, 67\penalty0 (2):\penalty0 301--320, 2005.

\end{thebibliography}

\end{document}